\documentclass[letterpaper,twocolumn,10pt]{article}
\usepackage{usenix2019_v3}

\usepackage{tikz}
\usepackage{amsmath}

\usepackage{filecontents}
\usepackage{listings}
\usepackage{color}
\usepackage{xspace}
\usepackage{graphicx}
\usepackage{tikz}
\usepackage{hyperref}
\usepackage{caption}
\usepackage{graphics}
\usepackage{array}
\usepackage{multirow}
\usepackage{amssymb}
\usepackage{subcaption}
\usepackage{pifont}
\usepackage{amsmath}
\usepackage{booktabs}
\usepackage{hhline}
\usepackage{pgfplots}
\usepackage[title]{appendix}
\usepackage{enumitem}

\newcommand{\eg}{\textit{e.g.,}\xspace}
\newcommand{\ie}{\textit{i.e.,}\xspace}
\newcommand{\rom}[1]{\textit{\expandafter\romannumeral #1}}

\newcommand{\meuzz}{{\sc Meuzz}\xspace}
\usepackage[singlelinecheck=on,font=small,labelfont=bf]{caption}
\usepackage[subtle]{savetrees}

\usepackage{cite}

\usepackage{url}

\makeatletter
\renewcommand\section{\@startsection{section}{1}{\z@}%
                       {-8\p@ \@plus -4\p@ \@minus -4\p@}%
                       {6\p@ \@plus 4\p@ \@minus 4\p@}%
                       {\normalfont\large\bfseries\boldmath
                        \rightskip=\z@ \@plus 8em\pretolerance=10000 }}
\renewcommand\subsection{\@startsection{subsection}{2}{\z@}%
                       {-8\p@ \@plus -4\p@ \@minus -4\p@}%
                       {6\p@ \@plus 4\p@ \@minus 4\p@}%
                       {\normalfont\normalsize\bfseries\boldmath
                        \rightskip=\z@ \@plus 8em\pretolerance=10000 }}
\renewcommand\subsubsection{\@startsection{subsubsection}{3}{\z@}%
                       {-4\p@ \@plus -4\p@ \@minus -4\p@}%
                       {-1.5em \@plus -0.22em \@minus -0.1em}%
                       {\normalfont\normalsize\bfseries\boldmath}}
\makeatother

\begin{document}

\date{}

\title{\Large \bf \meuzz: Smart Seed Scheduling for Hybrid Fuzzing}

\author{
{\rm Yaohui Chen}\\
Northeastern University \\
yaohway@gmail.com
\and
{\rm Mansour Ahmadi}\\
Northeastern University \\
Mansosec@gmail.com
\and
{\rm Reza Mirzazade farkhani}\\
Northeastern University \\
reza699@ccs.neu.edu
\and
{\rm Boyu Wang}\\
Stony Brook University \\
boywang@cs.stonybrook.edu
\and
{\rm Long Lu}\\
Northeastern University \\
l.lu@northeastern.edu
} %

\maketitle

\begin{abstract}
Seed scheduling highly impacts the yields of hybrid
fuzzing. Existing hybrid fuzzers schedule seeds based on fixed heuristics that
aim to predict input utilities. However, such heuristics are not generalizable
as there exists no one-size-fits-all rule applicable to different programs. They
may work well on the programs from which they were derived, but not others. 

To overcome this problem, we design a \underline{M}achine
learning-\underline{E}nhanced hybrid f\underline{UZZ}ing system (\meuzz), which
employs supervised machine learning for adaptive and generalizable seed
scheduling.
\meuzz determines which new seeds are expected to produce better fuzzing yields
based on the knowledge learned from past seed scheduling decisions made on the
same or similar programs.
\meuzz extracts  
a series of features for learning via code reachability and dynamic analysis, which
incurs negligible runtime overhead (in microseconds).  \meuzz
automatically infers the data labels by evaluating the fuzzing performance of
each selected seed. As a result, \meuzz is generally applicable to, and performs
well on, various kinds of programs.

Our evaluation shows \meuzz significantly outperforms the state-of-the-art
grey-box and hybrid fuzzers, achieving 27.1\% more code coverage than QSYM. The
learned models are reusable and transferable, which boosts fuzzing performance
by 7.1\% on average and improves 68\% of the 56 cross-program fuzzing campaigns.
When fuzzing 8 well-tested programs under
the same configurations as used in previous work,
\meuzz discovered 47 deeply hidden and previously unknown bugs, among which 21
were confirmed and fixed by the developers.

\end{abstract}

\section{Introduction}
\label{sec:intro}

Hybrid testing as a research topic has attracted tremendous attention and made
significant contributions to bug discovery. For instance, the winning teams in
the DARPA Cyber Grand Challenge\cite{cgc} all used hybrid
testing\cite{shellfish}. Compared with plain fuzzing, hybrid testing features an
extra concolic execution component, which revisits the fuzzed paths, solves the
path conditions, and tries to uncover new paths.

One key challenge in hybrid testing is to recognize high-utility seeds (\ie
seeds of high potential to guide concolic execution to crack complex conditions guarding more coverage and bugs).
Prioritizing such seeds allows the hybrid fuzzer to achieve higher code coverage
more quickly, and in turn, discover more bugs in a fixed time frame. Moreover,
this prioritization matters in practice because the concolic execution engine
usually has limited time budget and can explore only a (small) subset of all
fuzzer-generated seeds. Being able to estimate seed utility allows hybrid
fuzzers to use concolic execution more efficiently.

The existing work \cite{aflfast, afl, libfuzz, savior,qsyminsu,digfuzz,driller}
uses purely heuristic-based seed selection. For example, some prefer seeds with
smaller sizes while some value those that lead to new code coverage.
These heuristics, despite their simplicity, do not perform equally well across
different kinds of programs and are not universally suitable for all programs. 
Contradicting the previous belief~\cite{afl,libfuzz,qsymimpl}, our experiments show that seeds leading to
new coverage sometimes have the lowest utility (\S\ref{eval:insights}).
Similarly, 
previous work {\cite{afl,qsyminsu} suggested that smaller seeds should have higher
utility, which however is not true in certain programs as our evaluation shows.
As a result, these simple and fixed heuristics may cause non-optimum seed
selections, overwhelming the concolic engine with low-utility seeds and slowing
down bug discovery.

Compared to heuristics, Machine Learning (ML) algorithms, when trained with
sufficient data, can discover complex and implicit patterns automatically
\cite{mlrationalreasoning}. 
We show that seed selection strategies that are automatically learned based on
individual programs perform better than manually defined heuristics that fail to
consider all kinds of programs. As our experiment shows that the influence of each
feature varies across different programs, suggesting that no single feature (or
rule) can work well for all programs. 
ML-based seed selection avoids the need for manually designing, testing, and
reasoning about seed selection rules, which can be daunting, non-scalable, or
even impossible when the volume of data to be analyzed is overwhelming.

In this paper, we introduce \meuzz, an ML-enhanced hybrid fuzzing system. Unlike
existing work, which schedule seeds using simple heuristics derived from a
particular set of test programs, \meuzz uses ML and a set of static and dynamic
features, computed from seeds and individual programs, to predict seed utility
and perform seed scheduling. \meuzz also has a built-in evaluation module that
measures prediction quality for continuous learning and improvement. To the best
of our knowledge, \meuzz is the first work \cite{sandiasurvey} that applies ML
to seed prioritization and scheduling.

To effectively apply ML to seed scheduling for hybrid fuzzing, our design of
\meuzz pays special attention to two ML tasks: \emph{feature engineering} and
\emph{data labeling}. While these are the essential steps to bootstrap ML, they
could be time-consuming and thus too costly or impractical to be included in the
fuzzing workflow. For instance, feature extraction can be very slow if it
requires heavy computation or extensive data collection. Moreover, it is not
straightforward to quantify seed utility, which is essential for labeling.

To tackle the aforementioned challenges,  we first engineer a set of lightweight
features based on code reachability and dynamic analysis. Second, we propose a
labeling method using the input descendant tree to quantify the utility of a
seed. Our evaluation shows that \meuzz takes only 5${\mu}s$ on average to
extract an individual feature. It also confirms that the descendant tree of a
seed accurately reflects seed utility.

Collecting data and training a new model for every program might not be
economical or necessary. Therefore, we investigate the feasibility of model
reusability and transferability to answer the question: \emph{Is a learned model
transferable to different fuzzing configurations or programs?} 
Since the learning is designed to predict the likelihood of seeds triggering
bugs, rather than any specifics of the fuzzed program, a model learned by \meuzz
turns out to be applicable beyond the program from which the model is learned. 

We compare \meuzz with the state-of-the-art fuzzers~\cite{angora, afl, aflfast}
as well as the most recent hybrid testing systems\cite{savior,qsyminsu}. The results,
based on a set of real-world benchmark programs, show that \meuzz achieves much
higher code coverage than the tested fuzzers that use simple seed selection
heuristics. Particularly \meuzz expands the code coverage by as much as 27.1\%
compared to QSYM, the start-of-the-art hybrid fuzzing system. The experiments
also show that the prediction models learned by \meuzz have good reusability and
transferability. The reused models boost the coverage by 7.1\% on average. The
transplanted models improve fuzzing performance in 38 out of 56 cases (67.9\% of
cases), with 10 cases seeing more than 10\% improvement. 

This paper makes the following contributions.
\begin{itemize}
[topsep=0pt,itemsep=-1ex,partopsep=1ex,parsep=1ex]
\item {\textit{Effective and generalizable approach.}} We design, implement, and
evaluate  \meuzz, the first system that applies machine learning to the seed
selection stage of hybrid fuzzing. \meuzz performs better and is more widely
applicable than heuristic-based seed selection.

\item {\textit{Practical feature and label engineering.}} We address two major
challenges, namely feature engineering and label inference, when applying
ML to seed selection in hybrid fuzzing.
Our feature selection and extraction allow for online/continuous learning. They
are compatible with the existing hybrid fuzzing workflow and require no changes
to either fuzzers or concolic execution engines. We also propose an automatic
label inference method based on seed descendant trees.

\item {\textit{Reusable and transferable ML models.}} Our seed selection models
demonstrate strong {\it{reusability}} and {\it{transferability}}.
As a result, \meuzz can reuse a well-trained model on different programs (or
different fuzzing configurations) to quickly bootstrap the fuzzing campaign and
continuously improve and adapt the model to the current program or
configuration.  

\item {\textit{Open-Source.} The full implmentation of \meuzz will be open-sourced after acceptance.}
\end{itemize}

\section{Motivation}
\label{sec:motivation}

The seed selection (or scheduling) in fuzzing  aims to solve this problem: given
a program and a set of seeds, in which order the fuzzer should test the seeds to
maximize the gain during a fixed period. Seed selection plays a critical role in
hybrid fuzzing because the concolic execution engine can only explore an (often
small) subset of the seeds  due to time constraints. Hence, hybrid fuzzing
cannot fully benefit from concolic execution if the seed selection is not
optimal.

\smallskip
\noindent
\textbf{Why seed selection is important for hybrid fuzzing:}
Hybrid fuzzers without a seed scheduling mechanism (\eg Driller\cite{driller})
have to explore all inputs. This ``brute force'' strategy has two main
drawbacks.
First, concolic engines cannot keep up with the speed of plain fuzzing because
they run relatively slowly and often encounter path explosions and timeouts. As
an experiment, we used QSYM \cite{qsyminsu} to fuzz a set of real-world
benchmark programs. QSYM is one of the state-of-the-art concolic execution
engines for hybrid testing\footnote{Reportedly, QSYM is 3x faster than
Driller\cite{qsyminsu}.}. As shown in \autoref{fig:motiv}, for a continuous
24-hour run, QSYM was only able to explore 23.1\% of the seeds in fuzzer's
queue.

Second, a seed selection strategy affects fuzzing results drastically. A naive
strategy delays a fuzzer's exploration of interesting program locations, and
sometimes, prematurely forces the fuzzer to skip deep program paths and states.
Some recent research\cite{aflfast,digfuzz,savior,fuzzingscheduling,qsyminsu}
studied a few seed selection heuristics of various levels of sophistication. In
their experiments, fuzzers using these seed selection heuristics produce better
results (\eg higher code coverage) than fuzzers with naive or no strategies.

\begin{figure}[t]
    \resizebox{.40\textwidth}{!}{
        \centering
    \begin{tikzpicture}
\begin{axis}[
    ybar stacked,
	bar width=25pt,
	nodes near coords,
	every node near coord/.append style={xshift=0,yshift=-15pt,anchor=south,font=\footnotesize},
    enlarge x limits=0.2,
    ymin = -10,
    legend style={at={(0.8,0.98)},
      anchor=north,legend columns=1},
    ylabel={\# Seeds},
    symbolic x coords={tcpdump, libxml2, libjpeg, objdump},
    xtick=data,
    x tick label style={rotate=0,anchor=north},
    point meta=explicit symbolic,
    ]
\addplot[gray!20!black,fill=gray!70!white] plot coordinates {(tcpdump,6562) [6562] (libxml2,4355) [4355] 
  (libjpeg,3699) [3699] (objdump,2551) [2551]};
\addplot[ybar,gray!20!black,fill=gray!40!white] plot coordinates {(tcpdump,807)[807] (libxml2,1556) [1556]
  (libjpeg,1654) [1654] (objdump,832) [832]};
\legend{\strut Unexplored, \strut Explored}
\end{axis}
\end{tikzpicture}
}
    \caption{The total number of inputs explored by the concolic execution engine of QSYM in 24 hours. 
    On average, only 23.1\% of the inputs were explored by the concolic execution even though the engine was continuously running.}
    \vspace{-2ex}
    \label{fig:motiv}
\end{figure}
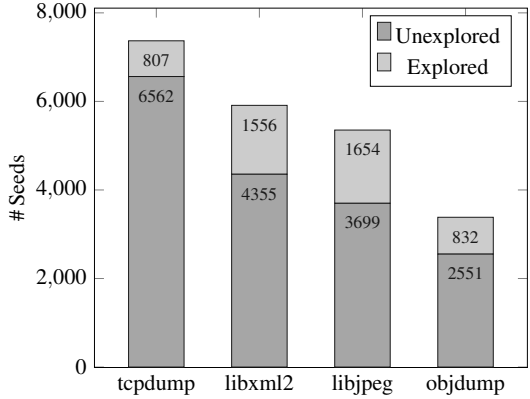

\smallskip
\noindent
\textbf{Why exploring machine learning for seed selection:}
All the existing seed selection strategies are based on manually defined
heuristics. Although performing well on their selected benchmarks, these
strategies may not be generalizable to, or suitable for, other programs. For
instance, DigFuzz\cite{digfuzz} and AFLFast\cite{aflfast} prioritize seeds with
less explored paths by fuzzer. Savior\cite{savior} prefers seeds dominating more
UBSan-labeled code paths. QSYM\cite{qsymimpl} prioritizes seeds with smaller
sizes. These heuristics are all based on intuition or empirical observations
gained from limited test cases or benchmarks.

A biased or unsuitable seed selection strategy delays or prevents fuzzers'
exploration of deep program states or the discovery of bugs. For instance,
QSYM\cite{qsymimpl} and ProFuzzer\cite{profuzzer} prioritize inputs with smaller
sizes. Their developers observed in their evaluation benchmarks that smaller
inputs lead to higher code coverage. However, as \cite{savior}
pointed out, QSYM fails to explore a large chunk of code in program {\tt{who}}
(a program in the LAVA-M benchmark\cite{lava})  due to the unsuitable seed
selection strategy (\ie only inputs larger than a certain size can trigger the
vulnerable functions in this case). 

This clearly indicates that fixed seed selection heuristics can hardly be
suitable for a wide set of programs (See \autoref{fig:modelchange}).

Due to the diverse scheduling scenarios, modern fuzzers (\eg AFL\cite{afltech}, QSYM\cite{qsymimpl}) often employ
multiple heuristics for seed prioritization. 
Unfortunately, relying on human efforts to learn and generalize seed selection
strategies, as the previous work did, is not scalable to a large number of features. 
In fact, it is just infeasible to manually reason about a big set of selection
criteria when the number of features and the amount of data to be analyzed become overwhelming (\eg OSS-Fuzz
generates four trillion seeds per week \cite{oss-fuzz} for different programs).

In contrast to heuristics, machine learning (ML) is good at discovering
underlying connections between data attributes \cite{mlrationalreasoning,
Mohri:2012:FML:2371238}. ML can be applied to seed selection because, as shown
by existing studies, the selection strategies are indeed learnable (\ie
exhibiting statistically significant patterns). With sufficient learning data, ML can not only infer the importance of different features but also mine the integration rules at scale.

\meuzz is the first to explore the ML-based, data-driven approach to
seed selection in hybrid fuzzing. Our result confirms that automatically and
continuously learned seed selection strategies are more suitable for individual
programs. 
\section{Background}
\label{sec:background}

Hybrid fuzzing \cite{driller,qsyminsu,savior} combines fuzzing and concolic execution
to address the deficiencies of both the approaches.
\autoref{fig:hybridfuzzing} shows an overview of a general hybrid fuzzing framework. The whole system
consists of three major components: fuzzer, concolic testing, and coordinator. For the sake of brevity,
we refer the interested readers to \cite{afltech,libfuzz,klee,sage} for the technical details of fuzzing and concolic 
execution. 

\subsection{Hybrid Fuzzing}
\begin{figure}[!ht]
    \centering
\includegraphics[scale=0.70]{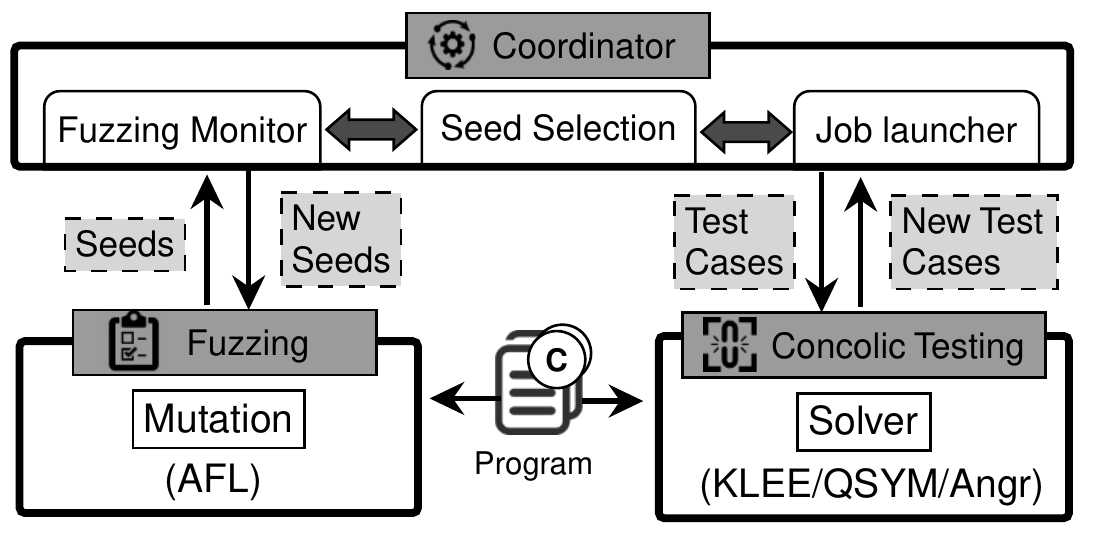}
\caption{General hybrid fuzzing workflow.}
\label{fig:hybridfuzzing}
\end{figure}

We dissect the coordinator component as it is less discussed in the literature and is the focus of this work.
The coordinator is a middleware that regulates the other two components. Its major tasks include
(\rom{1}) monitoring the fuzzer to decide when to launch the concolic execution engine, and (\rom{2})
prepare the running environments for concolic testing; and (\rom{3}) select and filter inputs that flow between
fuzzer and concolic executor. 

The seed selection module in the coordinator needs to decide which seeds in the fuzzer's queue
should be transferred to the concolic testing first (\ie Seed utility prediction phase). 
Before launching the concolic execution, the coordinator needs to rank all inputs in the fuzzer's queue based on their utility.
The \emph{utility} of seed should correspond to the estimation of its power to produce
additional coverage if it is selected to fuzz. As we mentioned in Section \ref{sec:motivation}, 
current methods use various heuristics to achieve this prioritization goal.

\subsection{Supervised Machine Learning}
Supervised ML is the task of learning from labeled data and applies the knowledge to unknown data. 
Classification and regression are two foremost categories of such algorithms.
While classification is used for predicting categorical responses,
regression predicts a numerical value to the new data based on previously observed data. 
Supervised learning has shown thriving employment in application security, 
including bug discovery \cite{CLEVER, DLIF, vuldeepecker}.

Supervised machine learning can be either online or offline. 
The difference between these two lies in how models are updated. 

\smallskip
\noindent
\textbf{Online learning:}
Some learning environments can change from second to second and their models need to 
get updated (or relearned) as fast as they see a new sample. Under this constraint of time, 
online learning shows promises by only considering the new data to update the model, 
which makes it an efficient approach. Basically, most learning algorithms that are compatible (but not limited) with the
standard optimization algorithms like stochastic gradient descent (SGD) can learn incrementally.

\smallskip
\noindent
\textbf{Offline learning:}
In contrast to online learning, the models in offline learning need to be retrained with the whole dataset as newer data appear. 
One of the successful examples of offline supervised learning techniques is Random Forest (RF), which has shown promising achievements, 
and in certain domains, has even better performance than neural networks \cite{hundredsclassifiers}. 
In addition to RF, deep learning has been shown success in different domains; however, 
they are usually practiced on unstructured data such as images and they require a relatively larger amount of data to 
perform well \cite{deeplearningvsrf}. Moreover, such techniques need high computational power and longer time to train; 
hence they are not suitable for the online fuzzing workflow.
\section{System Design}
\label{sec:design}

\subsection{System Overview}
\meuzz is the first machine learning-based hybrid fuzzer that learns
from the previously observed seeds
and identifies which kinds of seeds have the potentials to more effectively explore
the program being tested.

\autoref{fig:systemoverview} shows an overview of \meuzz.
\meuzz{} starts fuzzing (\ding{182}) a program with pre-defined or empty seeds.
It then extracts features (\ding{183}) from the program as well as the 
seeds (\S\ref{sec:featureengineering}) to model coverage gains. 
Such features are used to predict (\ding{184}) the coverage that unknown seeds may provide (\S\ref{sec:model}). Concolic engine (\ding{185}) then receives the potentially influential seeds from the prior step
and produces mutated seeds.
Next, \meuzz{} guides the fuzzer to use these seeds and their generated mutants--by the evolutionary algorithms--to continually test the program. 
In the beginning, the prediction model is randomly initialized, so the prediction quality is uncertain. 
But as fuzzing continues,  the model gets improved and will provide a more reliable prediction.
\meuzz{} updates the seed selection model in three steps. 
First, it infers the descendent trees (\ding{186}) of those seeds selected to the concolic engine in (\ding{185}); then, 
it derives a label (\ding{187}) based on the descendant trees of the previously selected seed (\S\ref{sec:seedlabelinference}); 
finally, it updates or retrains the model (\ding{188}) depending on the type of learning process (\S\ref{sec:model}, \S\ref{sec:updatingmodel}).

\begin{figure}[t]
\centering
\includegraphics[scale=0.6]{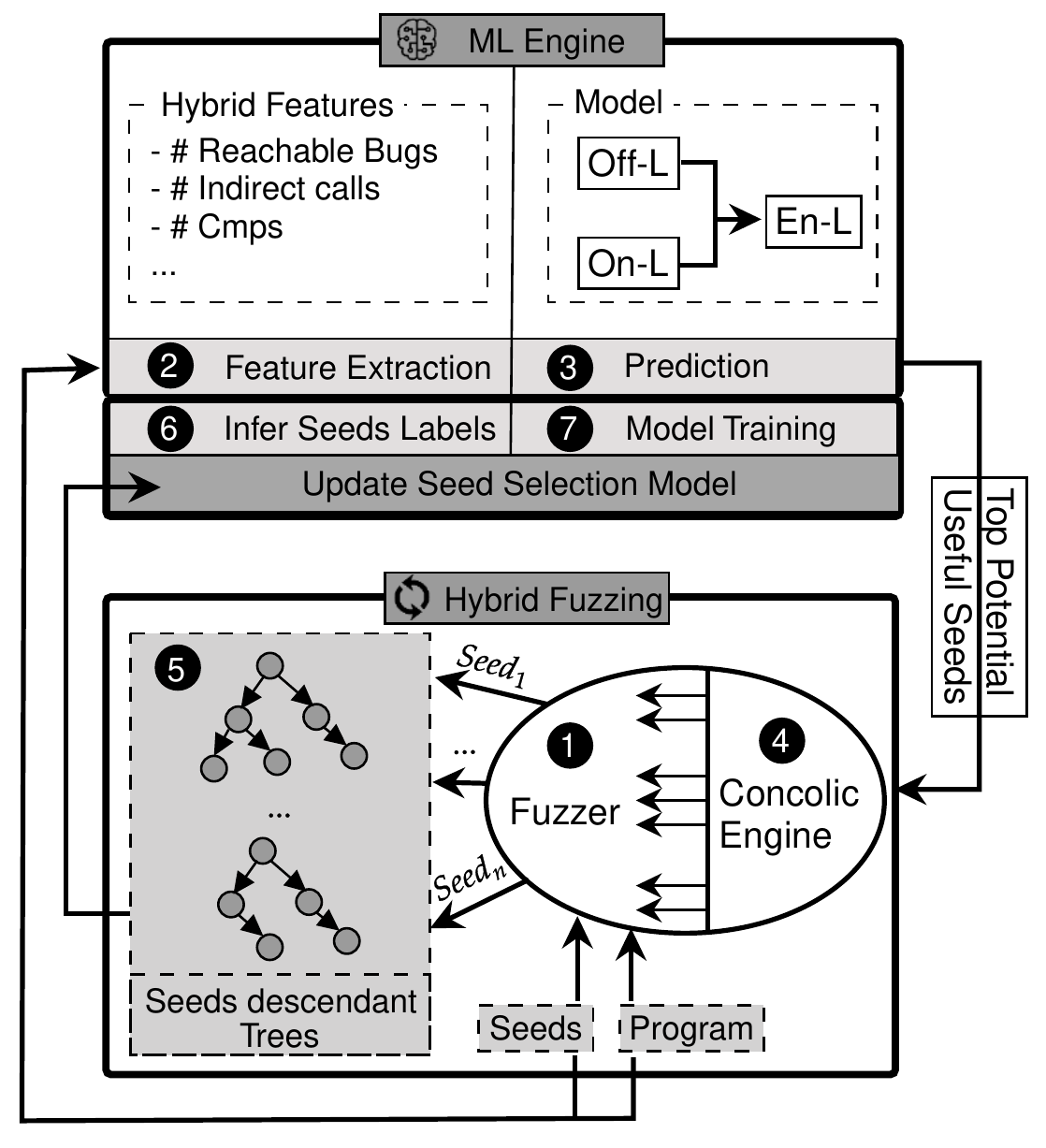}
\caption{System overview of MEUZZ. The coordinator is extended with a ML engine, which consists of 4 modules -- Feature extraction, label inference, prediction and training modules. During fuzzing, utility prediction and model training are carried out consecutively. After extracting features
	for inputs in the fuzzer's queue, the ML engine can predict their utilities based on the current model. Then, with the seed labels inferred from
	previously selected seeds, the model is trained iteratively with the new data.}
\label{fig:systemoverview}
\end{figure}

\subsection{System Requirements}
\label{sec:systemrequirements}

\meuzz aims to predict the seed utility in a more accurate and generalizable fashion than the
existing heuristic-based approaches while keeping the fuzzing efficiency intact. 
One of the steps that contribute the most in achieving these goals is 
feature extraction. \meuzz can potentially derive various semantic features because it has access to 
complex program structures, such as the Control Flow Graph (CFG) with sanitizer instrumentations. 
However, there are some challenges that 
\meuzz may encounter during feature extraction because it requires to adapt the ML engine to the 
online-style fuzzing workflow. To cope with such challenges, the feature engineering stage 
should meet the following requirements ($R1$--$R3$).

\smallskip
\noindent
\textbf{R1 - Utility Relevant:}
The ultimate goal of fuzzing is higher code coverage as well as discovering a higher number of hidden bugs.
The features should reflect the characteristics that may improve such measures.
For instance, how much a seed is likely to trigger more potential bugs or 
how much unexplored code a mutated seed will reach during its execution. 
As it is obvious, a seed is only meaningful in the context, 
which is the program it is executed upon. Accordingly, feature extraction needs to consider the seed and the program as
a bundle.

\smallskip
\noindent
\textbf{R2 - Seed-/Program-Agnostic:}
To achieve generalizability, the features should be seed-/program-agnostic. 
If a feature is target-dependent, it downgrades the ability to generalize. For example,
one could engineer a boolean feature based on the magic number that shows 
if a generated seed is genuine or not. Although this feature looks useful to ignore invalid seeds for fuzzing a specific program, 
it needs to be customized for fuzzing different programs as the inputs' formats change. 
Contrarily, ``meta properties''  like the execution path triggered by the input are more preferable, as it is a universally usable
characteristic regardless of the program.

\smallskip
\noindent
\textbf{R3 - Online Friendly:}
To keep the efficiency comparable to heuristic-based approaches, it is not only important 
how fast each feature can be extracted, 
but also the number of features is concerned during model construction. If the features are both light-weight and effective,
it is assured that the coordinator will not be blocked from launching the 
concolic executor and at the same time able to construct meaningful models to predict the seed utility. 
As a result, suitable features should strike a balance between analysis richness (\ie how informative is the analysis result) 
and computation complexity (\ie what is the time complexity for the analysis).

\subsection{Feature Engineering}
\label{sec:featureengineering}

The aforementioned requirements ($R1$--$R3$) guide us to engineer the following list of features. We
discuss them in four categories. 

\smallskip
\noindent
\textbf{Bug-triggering:}
Inspired by existing research\cite{savior}, we use the number of reachable sanitizer instrumentations as guidance for measuring 
how likely bugs can be triggered. As sanitizer instrumentations are based on sound analysis (\ie no missed bugs), 
it provides a good over-approximation when trying to quantify the number of bugs that can be found. 
Hence, we extract these two features:
\begin{enumerate}[topsep=2pt,itemsep=-1ex,partopsep=1ex,parsep=1ex]
	\item \emph{Count of reachable sanitizer instrumentations}: 
	For all branches throughout the path triggered by a given seed, 
	the number of reachable sanitizer instrumentations is computed and then sum up. 
	For instance, there are two branches in the left example of \autoref{fig:bugtriggeringfeatures}. 
	There are six potential bugs by following the branches, so the value for this feature is six.
	\item \emph{Count of reached sanitizer instrumentations}:
	For all branches throughout the path triggered by a given seed, we sum up the number of 
	reached sanitizer instrumentations by the fuzzer. The major difference between this feature and 
	the prior one is that this feature reflects the expectation of \emph{immediately} solvable sanitizer bugs, 
	while the former feature is an \emph{indirect} reflection. For instance, 
	the value of this feature in the right example of \autoref{fig:bugtriggeringfeatures} 
	is two because the potential bugs can be directly reached by negating the constraints from {\tt b1} and {\tt b2}.
\end{enumerate}

\begin{figure}[t]
    \centering   
    \includegraphics[width=0.48\textwidth]{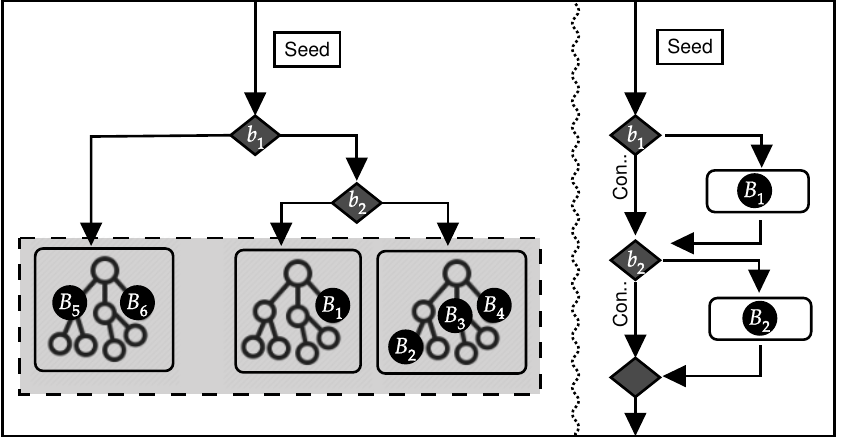}
    \caption{The examples that show how bug-triggering and coverage features are computed. }
    \vspace{-2ex}
    \label{fig:bugtriggeringfeatures}
\end{figure}
\setlength{\belowcaptionskip}{-4pt}

\smallskip
\noindent
\textbf{Coverage:}
Concolic execution is good at solving complex branch conditions. Hence if there are a lot of previously unsolved branches 
the concolic executor may encounter when executing on the given input, it will significantly improve the code coverage.
The most common situations where concolic execution can help is when a conditional statement
(\ie {\tt if-then-else} or {\tt switch-case}) exists. 
As the given input will only follow one of the branches, we call those branches stemmed from 
the same conditional statement \emph{neighbor branches}. So we extract the following feature to estimate each seed's potential of new coverage.

\begin{enumerate}[topsep=0pt,itemsep=-1ex,partopsep=1ex,parsep=1ex]
	\item \emph{Count of undiscovered neighbor branches}: For all branches along the path triggered by the given seed, we compare their neighbors, if any, with all previously triggered branches. We then sum up the previously undiscovered neighbors for each branch. 
	For instance,
	the value of this feature in the right example of \autoref{fig:bugtriggeringfeatures} is two if the
	seed follows the path with continue labels.
\end{enumerate}

\noindent
\textbf{Constraint Solving:}
We also devised a set of features that impact the solving capabilities of the concolic execution engine. 
The incentive behind selecting such features is that the performance of the
concolic executor significantly influences the entire hybrid fuzzing system.
\begin{enumerate}[topsep=2pt,itemsep=-1ex,partopsep=1ex,parsep=1ex]
	\item \emph{Count of external calls}: Existing concolic executors either rely on a simulated procedure or simply terminate the path execution when encountering an external function. As a result, external function calls may have negative impacts on the concolic executor, such as misleading the path and causing failure to generate correct seeds. 
    This feature records the count of external function calls along the path executed by the given seed.
	\item \emph{Count of comparison instructions}: This feature records the count of {\tt{cmp}} instructions 
    along the path executed by the given seed. Comparison instructions pose the constraints on the execution path, which will later
    be solved by the SMT solver. However, constraint-solving is very time-consuming and is often the reason for the timeout. 
     
	\item \emph{Count of indirect calls}: This is the number of indirect call instructions along the path executed by the given seed. 
    Indirect calls may cause state explosion because when the concolic executor encounters an indirect call with a symbolic pointer, it
    simply forks a state for each possible value that can be resolved for the symbolic pointer\cite{symexplain}. 
    In large programs, there could be many
    possible values for a symbolic function pointer.
	\item \emph{Length of path}: This feature records the number of executed branches (not deduplicated) by the given input.
    It helps identify the existence of large loops, which is another common reason that causes state explosion and solver timeout.
\end{enumerate}

\smallskip
\noindent
\textbf{Empirical:}
This set of features is devised based on the empirical observations by existing works. They might indirectly affect fuzzing performance.
\begin{enumerate}[topsep=2pt,itemsep=-1ex,partopsep=1ex,parsep=1ex]
	\item \emph{Input size}: Size of the input is often employed by existing tools as a heuristic to make a scheduling decision. 
	On the one hand, smaller size inputs often end the execution more quickly and then leave more time for the fuzzer or concolic executor 
	to explore other inputs\cite{qsymimpl,profuzzer}. On the other hand, larger input has a better chance to trigger 
	more functionalities\cite{savior}. Therefore, we consider the input size as one of the potential features for our approach.
	\item \emph{First seed with new coverage}: This is a boolean value indicating whether the given seed is the first one to 
	discover some new branches or not. This is based on the intuition that such seeds are more likely to trigger more new coverage. 
	This feature is used in many popular fuzzers\cite{afl,libfuzz}.
	\item \emph{Queue size}: This feature records how many inputs are saved in the fuzzing queue at the time of the query. If the queue is long, it is less likely to see more new coverage. Since \meuzz needs to predict the utility of each seed during runtime, namely how much \emph{more} new coverage can be discovered by fuzzing with the given input, the prediction should consider the current status of fuzzing.
\end{enumerate}

\subsection{Seed Label Inference}
\label{sec:seedlabelinference}

Labeling is an indispensable stage of data preprocessing in supervised learning. Well-defined labels make the prediction much easier and more reliable.
As we aim to predict the utility of a selected seed and there is no direct indication to show if the selected seed is definitely useful, we need to derive a label by which we show the proportion of the seed utility.

To understand the utility of a seed, we need to fuzz the program with that seed and check the outcome.
Fuzzers that use genetic algorithms (GAs) for seed generation represent such an outcome as a forest of \emph{input descendant tree},
which depicts the parent-child relationship of the seeds in the fuzzer's queue.
Each node of the tree represents a seed, and each edge connects a seed to one of its mutants.

In plain fuzzing, the root nodes are the original seeds provided by the user. 
Similarly, in hybrid testing, we model the inputs that are selected to be executed concolicly as the root nodes.
When an input is selected to explore, the concolic engine will produce mutants of the running input. 
These mutants can further cover the neighbor branches (\S~\ref{sec:featureengineering}) of the re-visited path. After these mutants get transferred back to the fuzzer's queue, the fuzzer can use GA to further
mutate them. As a result, we can draw the parent-child edges from the selected input to the mutants generated by the concolic engine, and to their GA-derived offsprings to form a mega descendant tree. 

If the descendant tree of a seed is larger, it comparatively means the seed 
contributes more to the fuzzer's code coverage. Hence, to derive the label, 
we measure the size of the input descendant tree of a seed and consider it as the label.

In reality, it is not feasible to compute the complete descendant tree since it could grow indefinitely 
if the user never terminates the fuzzing process.
As a result, we have to limit the tree analysis to a time window to make the label inference possible. 
Specifically, after the fuzzer imports
a seed from concolic executor, we wait for a certain number of fuzzing epochs for the fuzzer to explore the imported
seed and then compute the size of its descendant tree.

\subsection{Model Construction and Prediction}
\label{sec:model}

The next step after preparing the data is to predict the seed prominence (\ie label). 
As the seed labels are the number of nodes in the \textit{seed descendant tree}, 
their values are continuous so we need a regression model to predict them.
Hence, we embed a regression model in \meuzz{} in a way that when new seeds are generated by the fuzzer, 
the model predicts the utility of the seeds and then transfer the potential seeds to the concolic engine. 

\meuzz{} predicts very naively or just random at the beginning of fuzzing because the model just sees a few samples.
However, the prediction becomes more reliable when more seeds are generated--data plays a crucial role in advancing model--and the model receives updates.

As seeds are mutated continuously during fuzzing a program in real-time, prediction and
model update need to be done in a limited time window. Such limitation makes online learning approaches 
desirable candidates for model construction.
In online-learning, the model can be incrementally updated by only considering new data. 
It does not need to store all previous data and to learn a model from scratch in every iteration. 
Instead, the model can be updated incrementally based on the incoming input, previous model and 
historical fuzzing yields. Such an update is very fast and requires less storage, which fits our use case very well. 
Thus we adopt online learning as one of the techniques for model construction.

\subsection{Updating Model}
\label{sec:updatingmodel}

To assure the model is entirely up-to-date with the prevailing
seeds, ideally, we need to dynamically update/retrain the model, 
depending on the learning type (\ie online vs. offline).
By doing so, we can both predict and learn in real-time.

For online learning, we use the Recursive Least Square (RLS) algorithm\cite{shalev2014understanding, bottou1998online} to update our linear model. 
Suppose at time $t$, the input data and the label are $\mathbf{x}_t $ and $y_t$ correspondingly, 
where $\mathbf{x}_t$ is a vector of $d$-dimension. The following formula shows how the weight of the 
model at time $t$ (\ie $\mathbf{w}_{t}$) is updated based on the weight obtained from the previous model (\ie $\mathbf{w}_{t-1}$):
\begin{align*}
	\mathbf{w}_t = \mathbf{w}_{t-1} + \mathbf{C}_{t}^{-1} \mathbf{x}_t \left [y_t - \mathbf{x}_t^T \mathbf{w}_{t-1}  \right ]
\end{align*}
where $\mathbf{C}_{t}^{-1}$ is the inverse of for $\mathbf{C}_{t}$, and $\mathbf{C}_{t}$ is defined as:
\begin{align*}
\mathbf{C}_{t} = \sum_{i=1}^t\mathbf{x}_i \mathbf{x}_i +\lambda \mathbf{I} 
\end{align*}
Note that to calculate $\mathbf{C}_{t}^{-1}$, we do not need to store all previous data 
and compute the inverse. Based on the Woodbury formula, $\mathbf{C}_{t}^{-1}$ can also be 
updated recursively as follows:
\begin{align*}
\mathbf{C}_{t}^{-1} = \mathbf{C}_{t-1}^{-1} - \frac{\mathbf{C}_{t-1}^{-1} \mathbf{x}_t \mathbf{x}_t^T \mathbf{C}_{t-1}^{-1}}{1+ \mathbf{x}_t^T \mathbf{C}_{t-1}^{-1} \mathbf{x}_t}
\end{align*}
The complexity for such an update is $O(d^2)$.

To update the offline learning algorithms, the model needs to be retrained with all historical data in every iteration. 
Although retraining the model with the whole dataset every time a new seed is coming seems to be time-consuming, 
we show in our evaluation the approach is still practical in our case (\S\ref{eval:insights}). 
One reason is that the seed attributes are not of very high dimension and the number of seeds that need to be retrained is within an acceptable order of magnitude. 

\section{Implementation}
\label{sec:impl}

Among the three components of \meuzz, two of them are based on off-the-shelf software. We employ AFL-2.52b\cite{afl} for the fuzzing module and the re-engineered variant of KLEE from SAVIOR\cite{savior} for concolic execution.
We develop the coordinator component from scratch in Python in 3,152 SLOC.
Below, we detail the implementation of the major components of the ML engine, namely feature extraction and label inference.

\smallskip
\noindent
\textbf{Feature extraction:}
As discussed in \S\ref{sec:systemrequirements}, considering the trade-off between 
computational complexity and accuracy is key in feature extraction. 
Hence, for developing complicated features, we use a combination of static and dynamic analyses to offload the heavy tasks to compile time as much as possible. For instance, 
to extract the bug triggering features, we first instrument the target program with UBSan\cite{ubsanlist} at compile time. 
Then, a reachability analysis based on SVF\cite{svf} is used to extract the number of sanitizer instrumentations that can be reached from each branch. During runtime, we simply collect all 
the triggered branches by replaying the input and add up the number of reachable instrumentations from these branches.

To extract the feature of undiscovered neighbors, we record the branches and their neighbors at compile time.
This information is later used to query whether any neighbor of a triggered branch is covered. 
To facilitate fast queries, we
store the neighbor list as a disjoint-set data structure and use the union-find algorithm to query during runtime.

We extract the rest of the features either via compile-time instrumentation (\eg {\tt cmp}, {\tt call} instructions) 
and runtime input replay or via operating system APIs (\ie size, queue size, and new coverage).

\smallskip
\noindent
\textbf{Label inference:}
To collect the size of \textit{seed descendant tree}, we traverse AFL's fuzzing queue. 
Thanks to the seed naming system of AFL (\ie [{\tt id}, {\tt source}, {\tt mutation}, {\tt new cov}]), we can
iteratively traverse the seeds and use transitive closure to collect all the inputs imported from the concolic executor 
and their descendant trees.

\section{Evaluation and Analysis}
\label{sec:eval}
We conduct a comprehensive set of experiments to answer the following research questions:
\begin{itemize}
[topsep=0pt,itemsep=-1ex,partopsep=1ex,parsep=1ex]
    \item {\it RQ1}: Can ML-based seed scheduling outperform heuristics-based approaches (\S~\ref{eval:effectiveness} and \S~\ref{eval:bugs})?
    \item {\it RQ2}: Which features are more important in predicting seed utility and which learning mode is more effective (\S~\ref{eval:insights})?
    \item {\it RQ3}: Does the learned model adapt well to different fuzzing configurations (\S~\ref{eval:reusability})?
    \item {\it RQ4}: Is it feasible to transfer the learned model from a program to other programs to improve fuzzing yields (\S~\ref{eval:transferability})?
\end{itemize}

\subsection{Evaluation setup}
\definecolor{mygray}{gray}{0.9}

\begin{table}[t!]
\centering
\scriptsize
\caption{Evaluation settings}
\begin{tabular}{lcc|cc}
\toprule[0.5pt]
\toprule[0.5pt]

\multicolumn{3}{c|}{\bf{\emph{Program}}} & \multicolumn{2}{c}{\bf{\emph{Settings}}}   
 \\ \hline
 {\tt Name} & {\tt Version} & {\tt Driver} & {\tt Initial Seeds} & {\tt Options}
 \\ \hline

tcpdump & 4.10.0 & tcpdump & \cite{tcpdumpinput} & \tt{-r @@} \\

binutils & 2.32   & objdump & \cite{binutilinput}  & \tt{-D @@} \\

binutils & 2.32  & readelf &  \cite{binutilinput} & \tt{-A @@} \\

libxml & 2.9.9 & xmllint & \cite{libxmlinput} &  \tt{stdin} \\

libtiff & 4.0.10 & tiff2pdf& \cite{libtiffinput} & \tt{@@} \\

libtiff & 4.0.10 & tiff2ps & \cite{libtiffinput} & \tt{@@} \\

jasper & 2.0.16 & jasper & ~\cite{libjpeginput} & \tt{-f @@ -T pnm} \\

libjpeg & jpeg9c & djpeg & ~\cite{libjpeginput} &  \tt{stdin} \\

\bottomrule[0.5pt]
\bottomrule[0.5pt]
\end{tabular}

\label{tab:eval-setup}
\vspace{-2ex}
\end{table}

Following the general fuzzing evaluation guideline\cite{evaluatefuzz}, we choose 8 real-world benchmark programs commonly used by existing work \cite{savior,qsyminsu,digfuzz,aflfast,angora}.
Table~\ref{tab:eval-setup} shows the configurations used for fuzzing each program. 
All experiments are conducted on AWS c5.18xlarge servers running Ubuntu 16.04 with 72 cores and 281 GB RAM. 
Without explicitly mention, all tests run for 24 hours each by assigning three CPU cores to each fuzzer and are repeated at least 5 times; we report the average result with Mann-Whitney U-test.  

We compare \meuzz with the state-of-the-art grey-box fuzzers,
such as AFL\cite{afl}, AFLFast\cite{aflfast}, and Angora\cite{angora}, as well
as hybrid testing systems including QSYM\cite{qsyminsu} and SAVIOR\cite{savior}.
The seed selection modules of all these previous systems are based on heuristics.
We could not test Driller \cite{driller} on the chosen benchmarks because its
concolic execution engine fails to run them. Moreover, we test Vuzzer
\cite{vuzzer} and T-Fuzz\cite{tfuzz} but we compare them with \meuzz in a
different way than we do with the other fuzzers. This separate comparison is
because these two fuzzers do not
support concurrent fuzzing. Due to
the space limit, we discuss our observations and show the results of their branch
coverage in Appendix~\ref{sec:appendix:other}.

For \meuzz, we consider three different
configurations according to the learning process, namely \meuzz-OL, \meuzz-RF and \meuzz-EN, which refer to
 online learning linear model, offline learning random forest model and the arithmetic average of the first two models' utility predictions, respectively.

Since Savior and QSYM need at least three CPU cores, we enforce this fuzzing setting to all the fuzzers to build a fair comparison environment.
We launch one master and two slaves for the grey-box fuzzers; and
one master, one slave, and one concolic execution engine for the hybrid fuzzers. To reduce the randomness of OS scheduling, we pin each component of the fuzzers on the specific core.
Because \meuzz and SAVIOR instrument the testing program with UBSAN\cite{ubsanlist}, 
we also apply this sanitizer to all other fuzzers, as enabling sanitizers is shown to
improve the fuzzer's effectiveness for finding bugs.

\subsection{Learning Effectiveness}
\label{eval:effectiveness}
\setlength{\belowcaptionskip}{0pt}
\begin{figure}[!ht]
    \centering   
    \begin{subfigure}[b]{0.24\textwidth}
        \centering
        \includegraphics[width=1\textwidth]{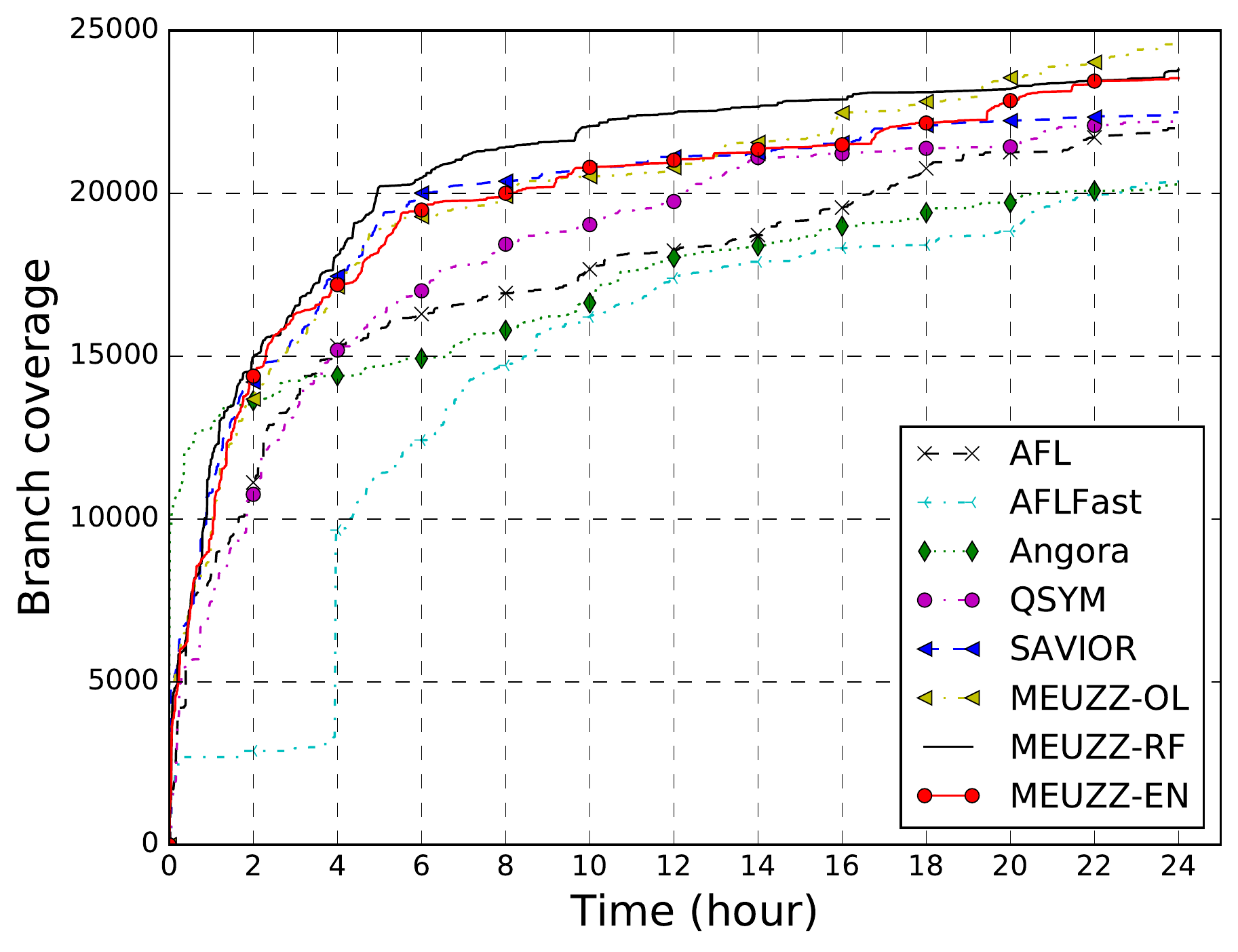}
        \vspace*{-15pt}
        \caption{\scriptsize{tcpdump branch coverage ($p_{1}$=$0.071$, $p_{2}$=$0.005$, $p_{3}$=$0.082$)}}
        \label{fig:cov:tcpdump}
    \end{subfigure}
        \begin{subfigure}[b]{0.23\textwidth}
        \centering
        \includegraphics[width=1\textwidth]{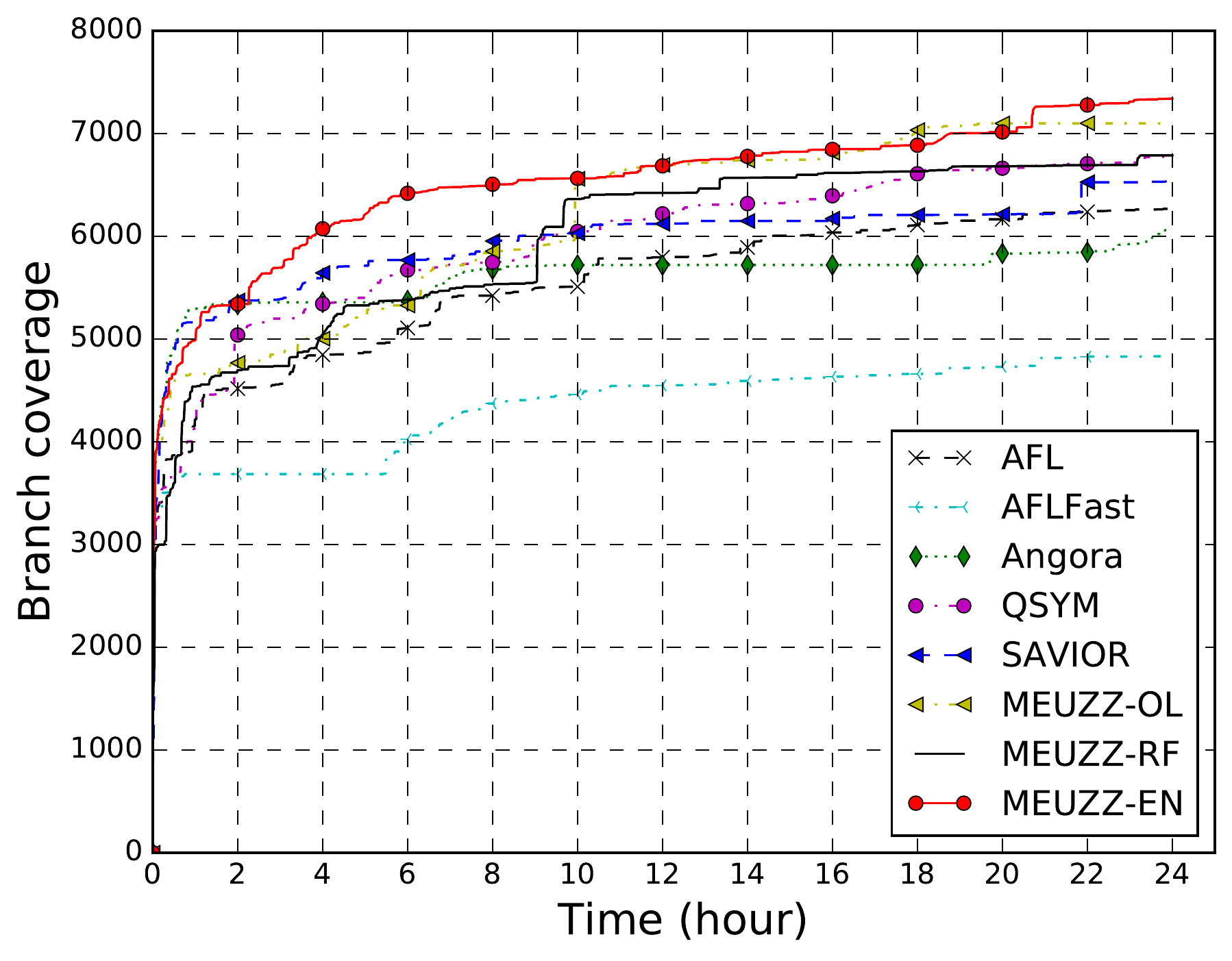}
         \vspace*{-15pt}
        \caption{\scriptsize{objdump branch coverage ($p_{1}$=$0.044$, $p_{2}$=$0.056$, $p_{3}$=$8.2*e^{-4}$)}}
        \label{fig:cov:objdump}
    \end{subfigure}\\
    \begin{subfigure}[b]{0.23\textwidth}
        \centering
        \includegraphics[width=1\textwidth]{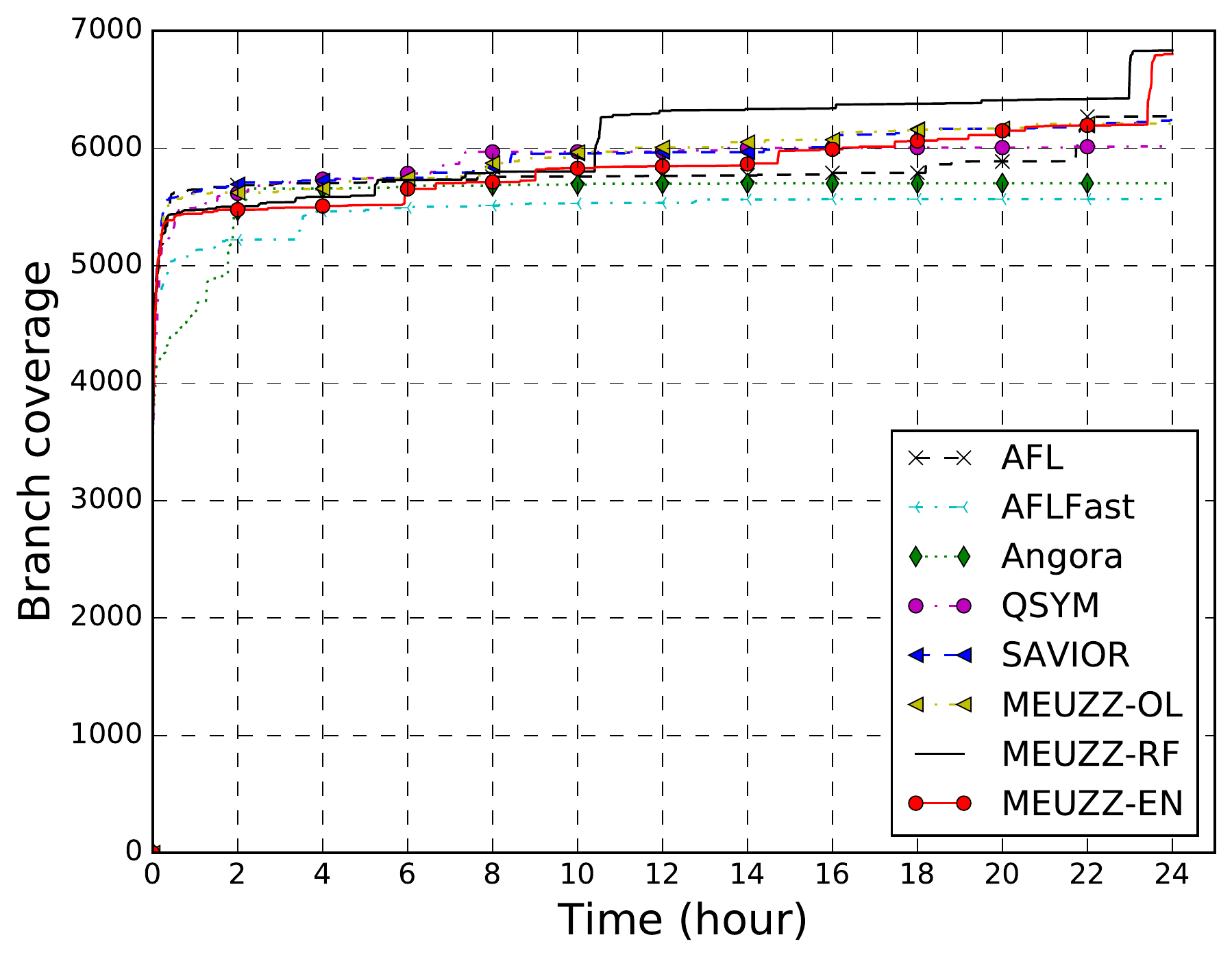}
         \vspace*{-15pt}
        \caption{\scriptsize{libxml branch coverage ($p_{1}$=$0.035$, $p_{2}$=$0.059$, $p_{3}$=$0.054$)}}
        \label{fig:cov:libxml}
    \end{subfigure}
    \begin{subfigure}[b]{0.23\textwidth}
        \centering
        \includegraphics[width=1\textwidth]{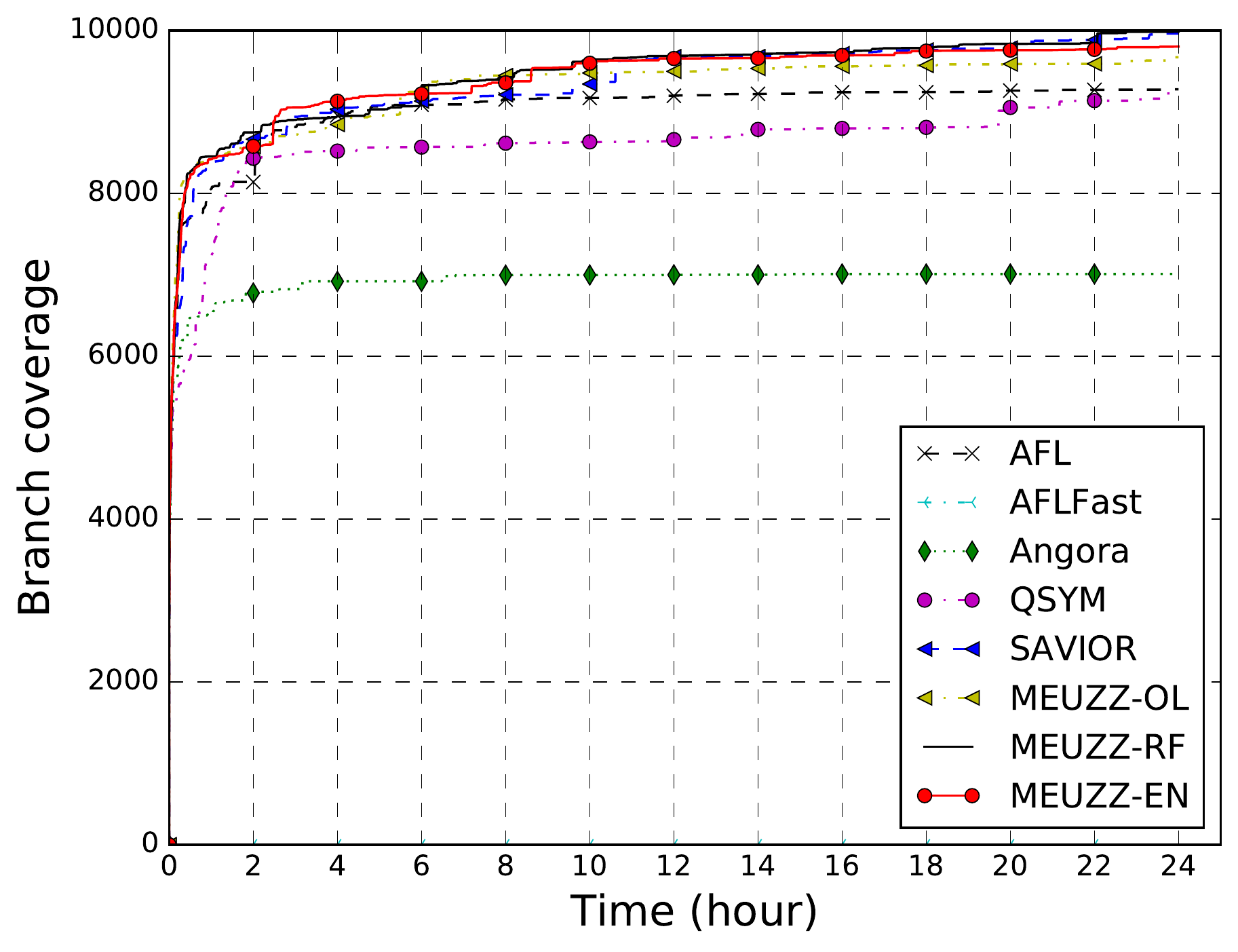}
       \vspace*{-15pt}
        \caption{\scriptsize{tiff2pdf branch coverage ($p_{1}$=$8.2*e^{-4}$, $p_{2}$=$5.6*e^{-4}$, $p_{3}$=$6.2*e^{-5}$)}}
        \label{fig:cov:tiff2pdf}
    \end{subfigure}\\
    \begin{subfigure}[b]{0.23\textwidth}
        \centering
        \includegraphics[width=1\textwidth]{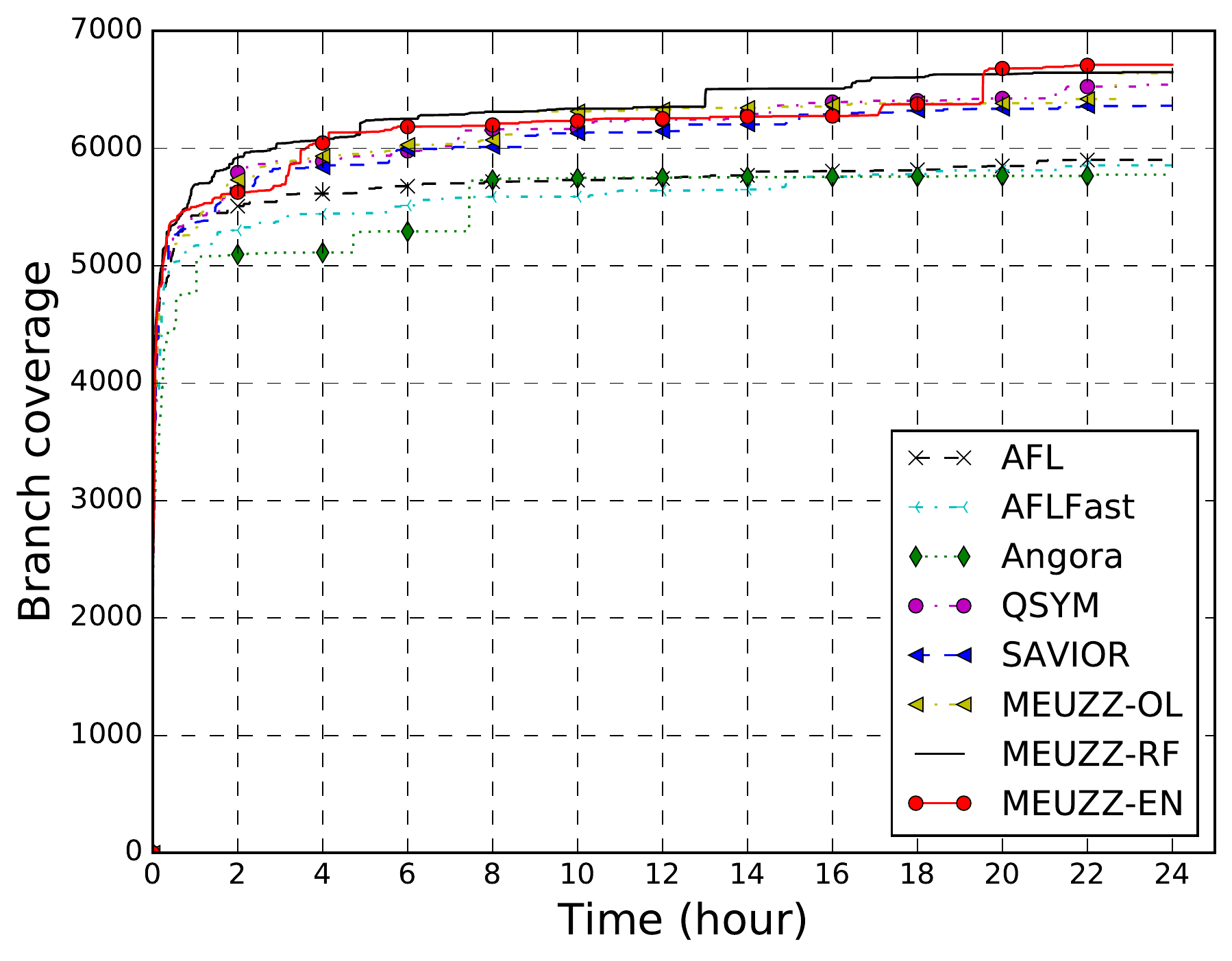}
        \vspace*{-15pt}
        \caption{\scriptsize{tiff2ps branch coverage ($p_{1}$=$0.035$, $p_{2}$=$0.091$, $p_{3}$=$0.017$)}}
        \label{fig:cov:tiff2ps}
    \end{subfigure}  
    \begin{subfigure}[b]{0.23\textwidth}
        \centering
        \includegraphics[width=1\textwidth]{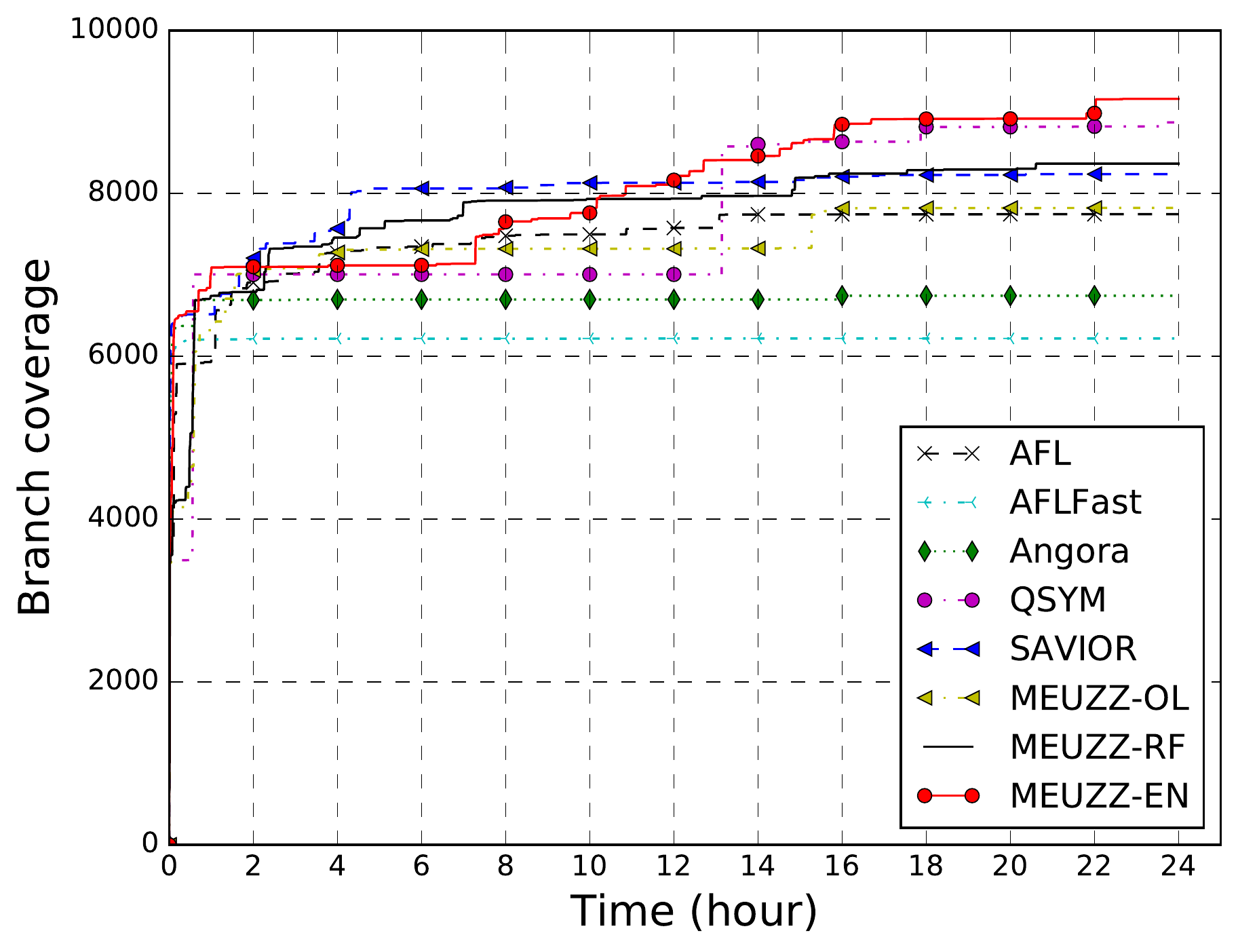}
       \vspace*{-15pt}
        \caption{\scriptsize{jasper branch coverage ($p_{1}$=$0.037$, $p_{2}$=$0.192$, $p_{3}$=$0.015$)}}
        \label{fig:cov:jasper}
    \end{subfigure}\\
    \begin{subfigure}[b]{0.23\textwidth}
        \centering
        \includegraphics[width=1\textwidth]{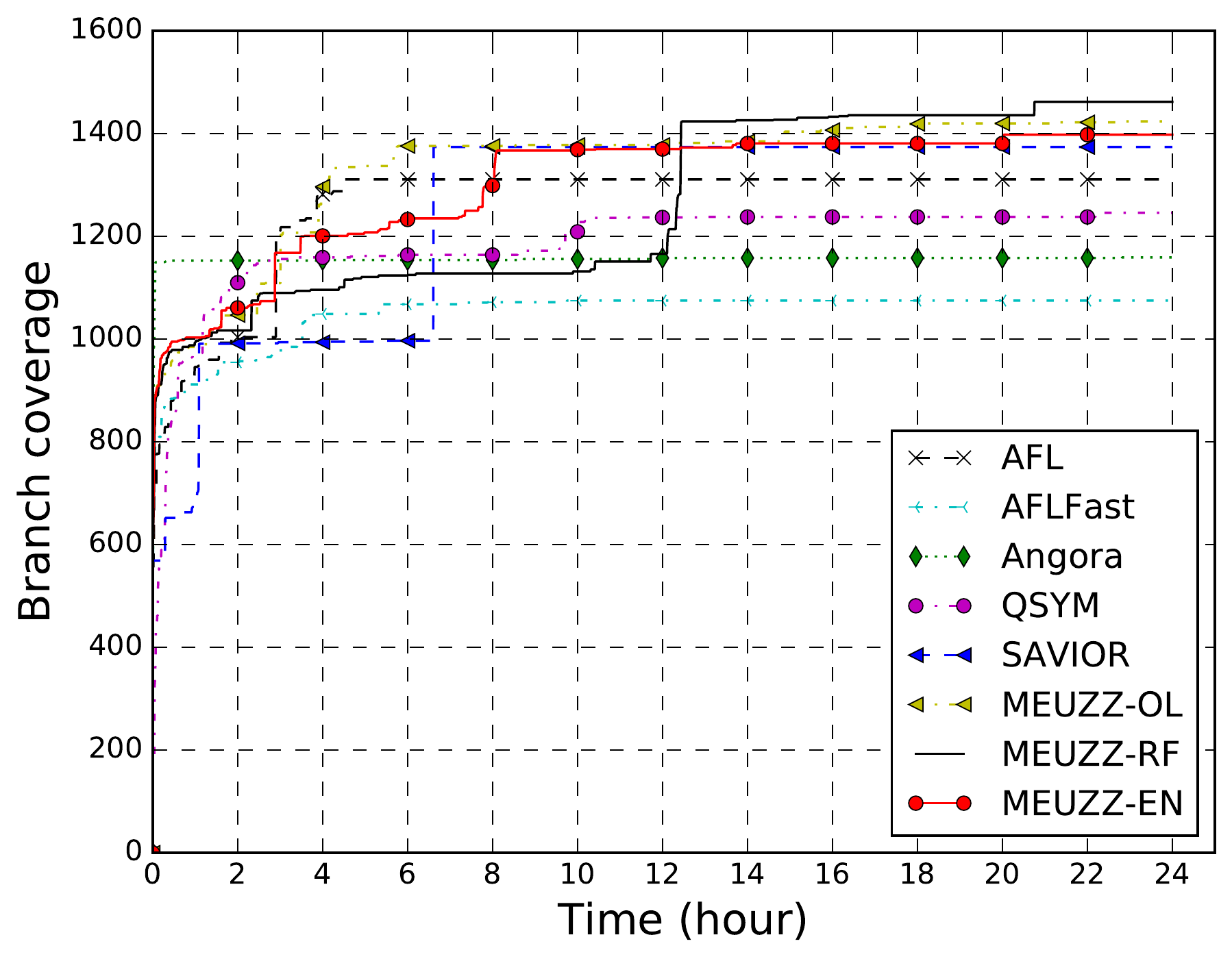}
       \vspace*{-15pt}
        \caption{\scriptsize{readelf branch coverage ($p_{1}$=$0.012$, $p_{2}$=$0.093$, $p_{3}$=$8.2*e^{-4}$)}}
        \label{fig:cov:readelf}
    \end{subfigure}
    \begin{subfigure}[b]{0.23\textwidth}
        \centering
        \includegraphics[width=1\textwidth]{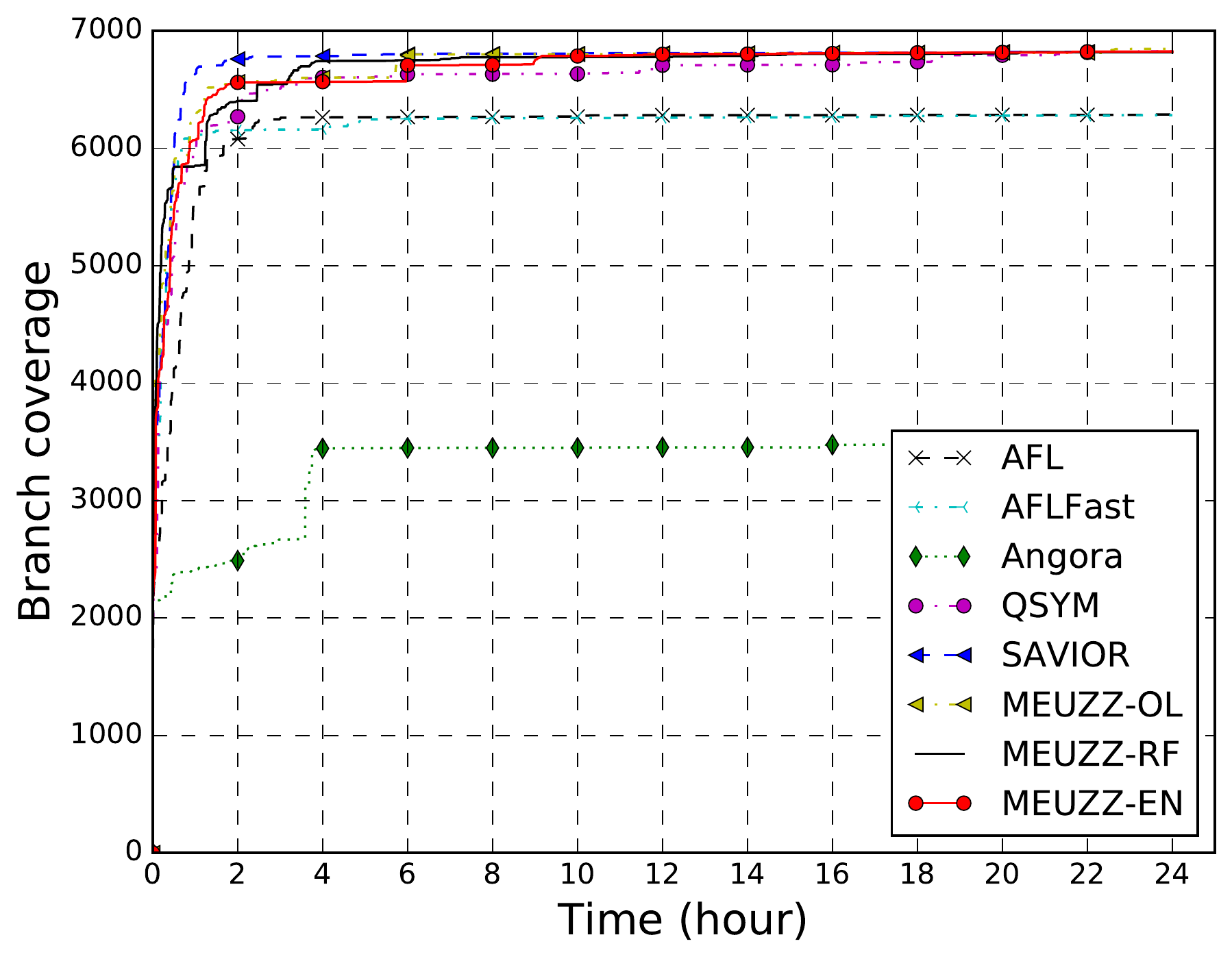}
       \vspace*{-15pt}
        \caption{\scriptsize{djpeg branch coverage ($p_{1}$=$0.072$, $p_{2}$=$0.021$, $p_{3}$=$0.093$)}}
        \label{fig:cov:djpeg}
    \end{subfigure}
    \caption{Branch coverage fuzzing with valid seeds (higher is better). $p_{1}$, $p_{2}$ and $p_{3}$ are p-values in Mann-Whitney U Test by comparing QSYM with MEUZZ-OL, MEUZZ-RF and MEUZZ-EN, respectively.}
    \label{fig:codecov}
\end{figure}

The most straightforward metric to measure the effectiveness of 
\meuzz is code coverage, which is also a widely accepted and evaluated metric.
\autoref{fig:codecov} shows the branch coverage achieved by different fuzzers to the required time for fuzzing. 
Based on the coverage result, we have several interesting findings.

First, \meuzz covers more code than other fuzzers in most programs after 24 hours of fuzzing. Among the \emph{non-ML} fuzzers, 
QSYM performs the best in terms of code coverage, thanks to its efficient concolic execution engine tailored specially for 
hybrid fuzzing.
Compared with QSYM, the \meuzz variants achieve various levels of coverage improvements. In {\tt{tcpdump}}, {\tt{objdump}}, 
{\tt{readelf}} and {\tt{libxml}}, \meuzz improves code coverage over QSYM by more than 10\%, and particularly 27.1\% by \meuzz-RF
in {\tt readelf}. In {\tt tiff2pdf} and {\tt tiff2ps}, \meuzz also has moderate coverage improvements. However, in {\tt jasper} 
and {\tt djpeg}, there is no much difference between \meuzz and QSYM; we speculate it is because all fuzzers are saturated
and hit a plateau after 6 hours.

Second, \meuzz covers less code in the beginning but gradually surpasses other fuzzers as time progresses. 
For example, in {\tt objdump} \meuzz-OL and \meuzz-RF did not cross QSYM and SAVIOR until after 9.6 hours of fuzzing, but \meuzz eventually 
achieves 14\% higher code coverage. Similar situations can be observed in {\tt libxml}, {\tt readelf} and {\tt tiff2ps}. 
This observation is expected, as \meuzz starts seed scheduling with random parameters, hence the
performance of seed selection is unpredictable at the beginning. But as time passes, fuzzing data are increasingly collected and used to
refine the prediction model. Hence, the prediction becomes more accurate.

Lastly, the effectiveness of ML is presented in \autoref{fig:modelchange} in Appendix~\ref{sec:appendix:feature}. It is shown that
different programs are variously affected by different sets of features. 
For instance, \emph{External Calls} has more influence on six of the programs except for {\tt tcpdump} and {\tt djpeg}, showing that 
no single feature is sufficient to predict high-utility seeds. By using a data-driven
approach, we cannot only \emph{automatically} select the high
impactful features in different programs or situations, but also integrate them in a more optimal way than manual-crafting rules.

\subsection{Insights and Analyses}
\label{eval:insights}
\begin{figure}[ht]
\centering
\includegraphics[scale=0.5]{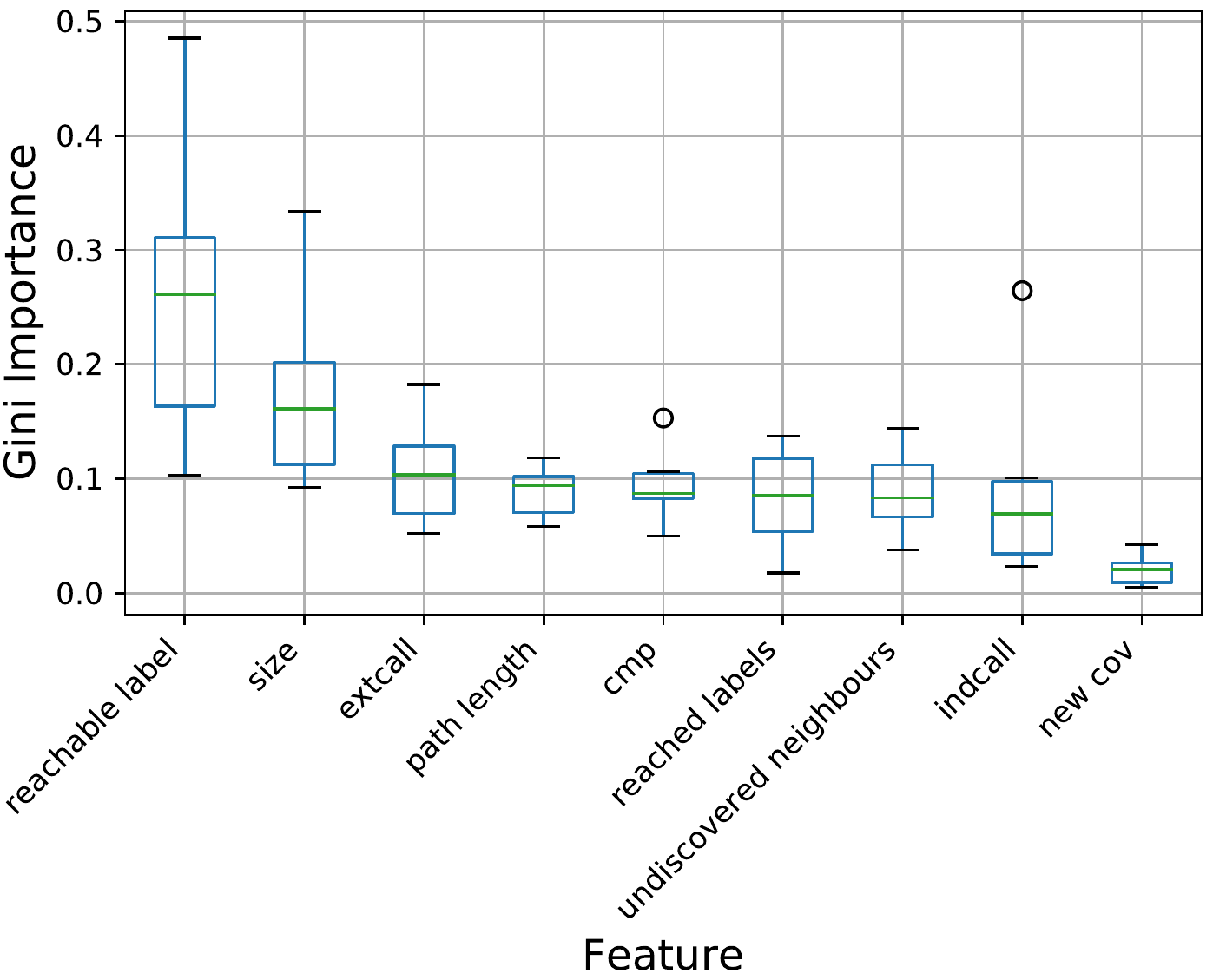}
\caption{The box plots show the importance of the features on nine programs. The importance is extracted by training an offline random forest model and they are ranked by the median of their importance. \textit{Reachable label} and \textit{New Cov} are the most and the least important ones, respectively.}
\label{fig:featureimportance}
\end{figure}

\begin{table}[t]
 \caption {Execution time spend on different learning stages}
 \resizebox{0.49\textwidth}{!}{
     \begin{tabular}{cc|cc|c}
     \toprule
     \multicolumn{2}{c}{Model Update (\tt{s})} & \multicolumn{2}{c}{Prediction(\tt{s})} & \multirow{2}{*}{Feature Extraction (\tt{s})} \\
     Online & Offline & Online & Offline & \\
     \hhline{=====}
          0.000636     & 0.326139 & 0.000016     & 0.003168  & $5\mathrm{e}{-6}$    \\
     \bottomrule
     \end{tabular}
     }
     \label{tab:learningtime}
\end{table}

\smallskip
\noindent
\textbf{Online v.s Offline learning:}
As mentioned in the previous section, offline learning with the random forest model sometimes beats online learning with the linear model; 
however, the main concern with using offline learning is time delays, especially during the model updating stage.

To further analyze the effects of time delays caused by offline learning, we profile each learning
stage during the 24 hours of fuzzing and report the average time spend on different learning steps. 
As shown in Table~\ref{tab:learningtime}, although offline learning spent 512x and 198x more time 
than online learning on updating the model and making
predictions, respectively, the absolute time-lapse is negligible (\ie in milliseconds). 
Hence, offline learning is not a critical hindrance throughout the hybrid fuzzing loop, which endorses the offline learning effectiveness discussed in Section \ref{eval:effectiveness}. 
Having said that, if fuzzing continues for a longer time and the number of seeds significantly increases, offline learning can become an obstacle.

\smallskip
\noindent
\textbf{Feature Analysis:}
\autoref{fig:featureimportance} presents the distribution of the importance of each feature separately in all programs.
The importance score is computed by capturing the \textit{mean decrease impurity} from the offline random forest models \cite{breiman1984classification}.
The figure shows the contribution of the \textit{New Cov} feature is the least among all the features. While it is difficult to entirely disregard the minor contribution of \textit{New Cov}, 
this suggests that putting much effort to follow the seeds that bring new coverage might jeopardize the chance to explore unknown seeds. 
This is also known as the famous Multi-Armed Bandit (MAB) problem\cite{mab}. 
This finding might shed some light on the
scheduling algorithm implemented in the popular fuzzers like AFL\cite{afl} that heavily rely on the \textit{New Cov} heuristics.

Also, the variance of change in the figure shows some of the features like \textit{Path Length} and \textit{New Cov} are less subject to programs, 
while others like \textit{Reachable Label} are more tied to programs. 
If the extraction of a feature heavily depends on static analysis, it is less precise compared with dynamic analysis 
because the sensitivity of static analysis affects the precision (\ie flow/context/field sensitivity).
We speculate this is one of the reasons that make a feature (\eg \textit{Reachable Label}) more dependent on individual programs. 
Also, there are additional factors that might affect dependability.
Program loops as a common trait in all programs uniformly affect the \textit{Path Length} feature, which makes the feature more agnostic to programs. 
Similarly, \textit{New Cov} is set to a seed
during runtime when it is the first one to trigger new behaviors (\eg coverage); this attribute is 
generally applicable to a variety set of programs.

It is worth noting that the average time 
to extract each feature is only 5${\mu}s$ (as shown in Table~\ref{tab:learningtime}),
thanks to our light-weight feature extractions.
This indicates that the online-friendly requirement is satisfied in \meuzz.

\subsection{Model Reusability}
\label{eval:reusability}

\setlength{\belowcaptionskip}{0pt}
\begin{figure}[!ht]
    \centering   
    \begin{subfigure}[b]{0.23\textwidth}
        \centering
        \includegraphics[width=1\textwidth]{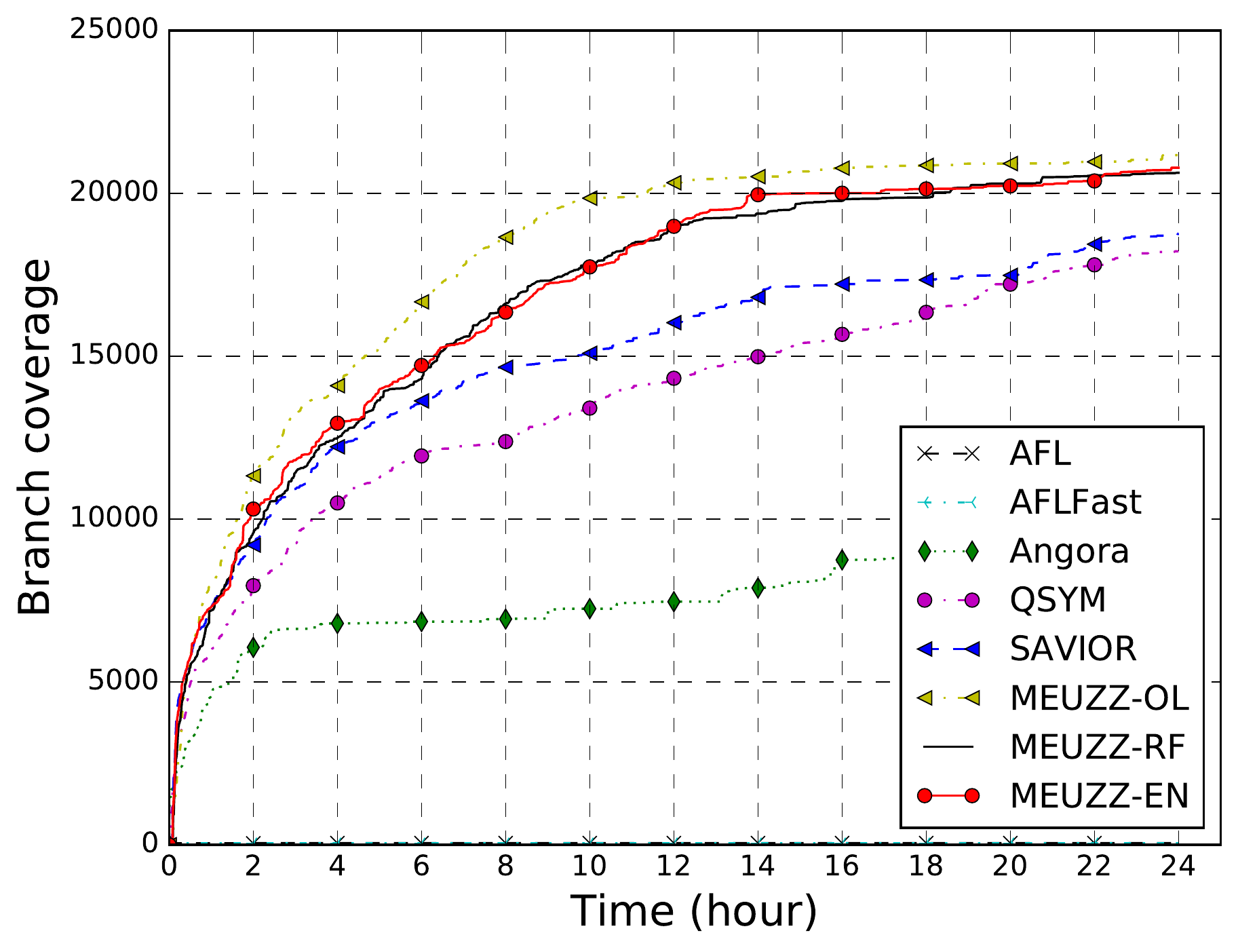}
        \vspace*{-15pt}
        \caption{\scriptsize{tcpdump branch coverage ($p_{1}$=$0.047$, $p_{2}$=$0.018$, $p_{3}$=$0.026$)}}
        \label{fig:cov2:tcpdump:naive}
    \end{subfigure}
        \begin{subfigure}[b]{0.23\textwidth}
        \centering
        \includegraphics[width=1\textwidth]{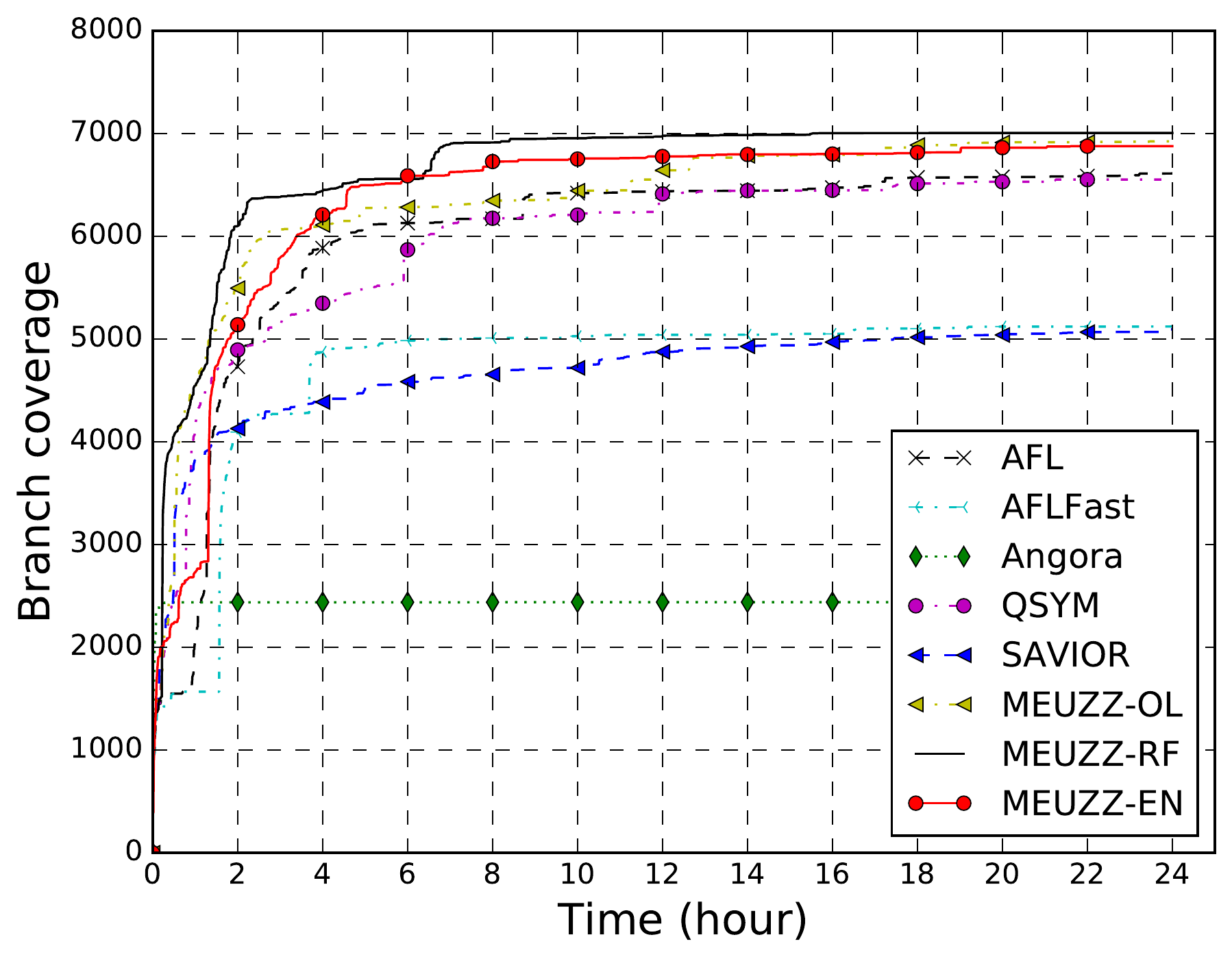}
         \vspace*{-15pt}
        \caption{\scriptsize{objdump branch coverage ($p_{1}$=$0.051$, $p_{2}$=$2.33*e^{-3}$, $p_{3}$=$5.7*e^{-3}$)}}
        \label{fig:cov2:objdump}
    \end{subfigure}\\
    \begin{subfigure}[b]{0.23\textwidth}
        \centering
        \includegraphics[width=1\textwidth]{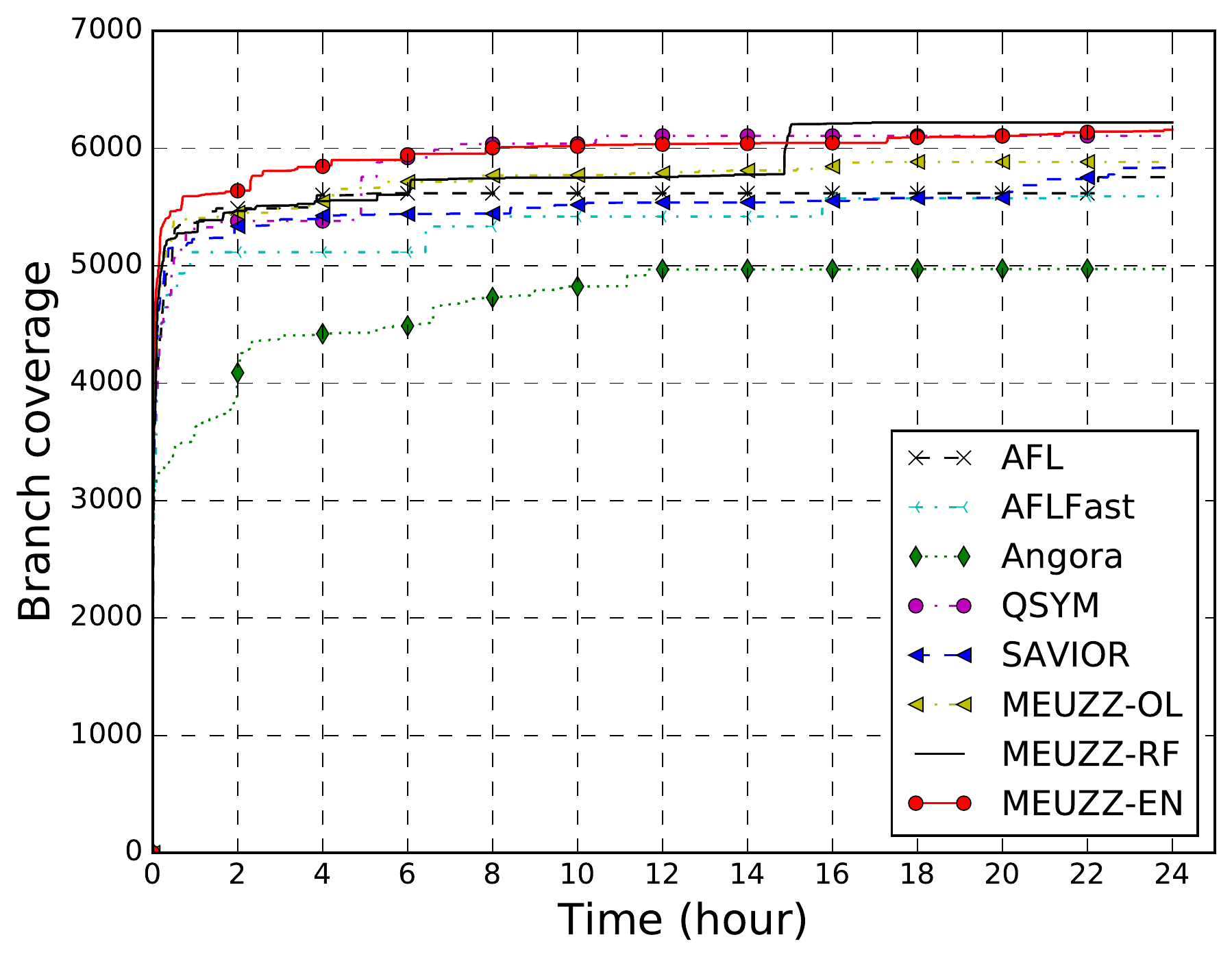}
         \vspace*{-15pt}
        \caption{\scriptsize{libxml branch coverage ($p_{1}$=$0.072$, $p_{2}$=$0.032$, $p_{3}$=$0.026$)}}
        \label{fig:cov2:libxml}
    \end{subfigure}
    \begin{subfigure}[b]{0.23\textwidth}
        \centering
        \includegraphics[width=1\textwidth]{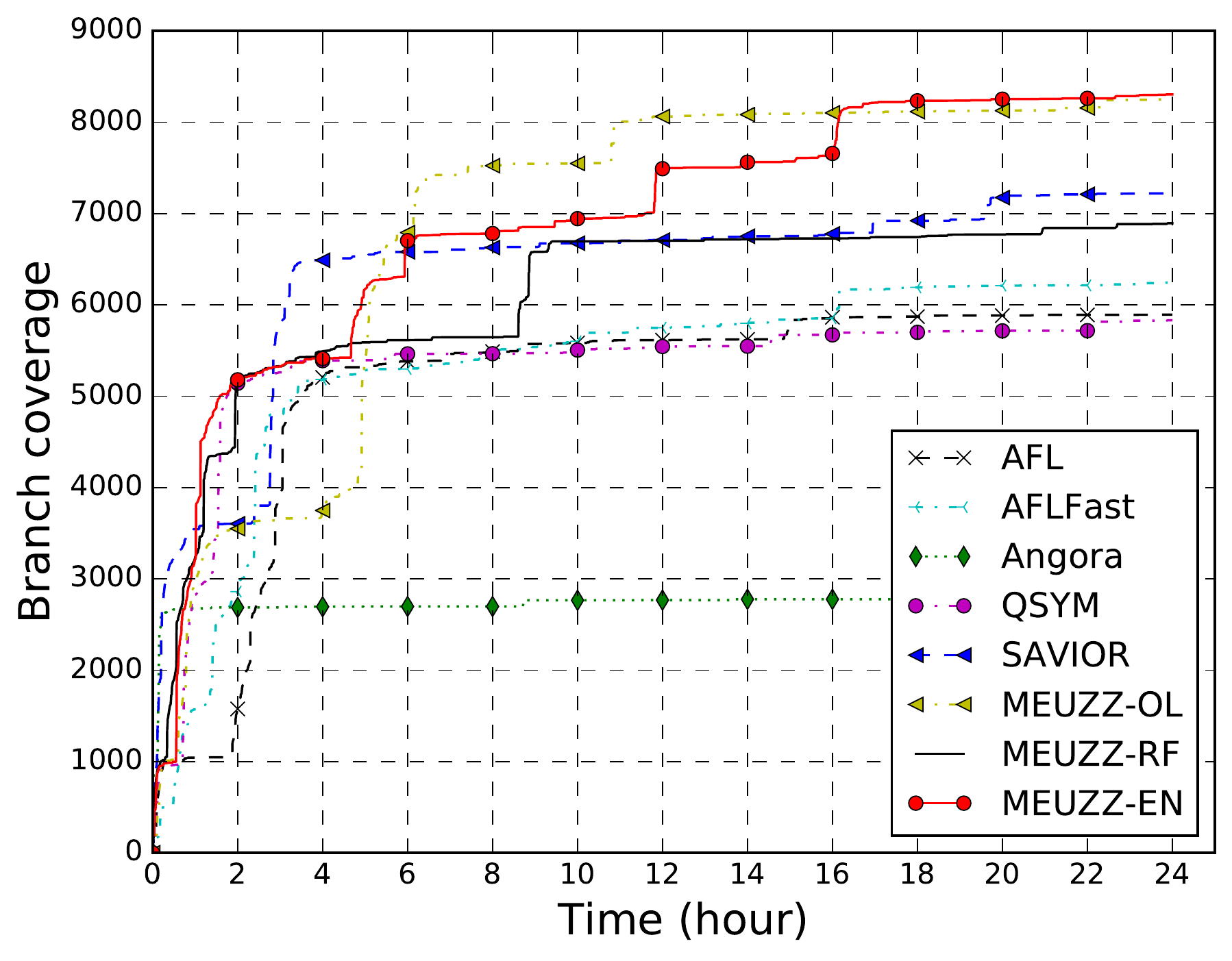}
       \vspace*{-15pt}
        \caption{\scriptsize{tiff2pdf branch coverage ($p_{1}$=$0.02$, $p_{2}$=$0.03754$, $p_{3}$=$5.7*e^{-3}$)}}
        \label{fig:cov2:tiff2pdf}
    \end{subfigure}\\
    \begin{subfigure}[b]{0.23\textwidth}
        \centering
        \includegraphics[width=1\textwidth]{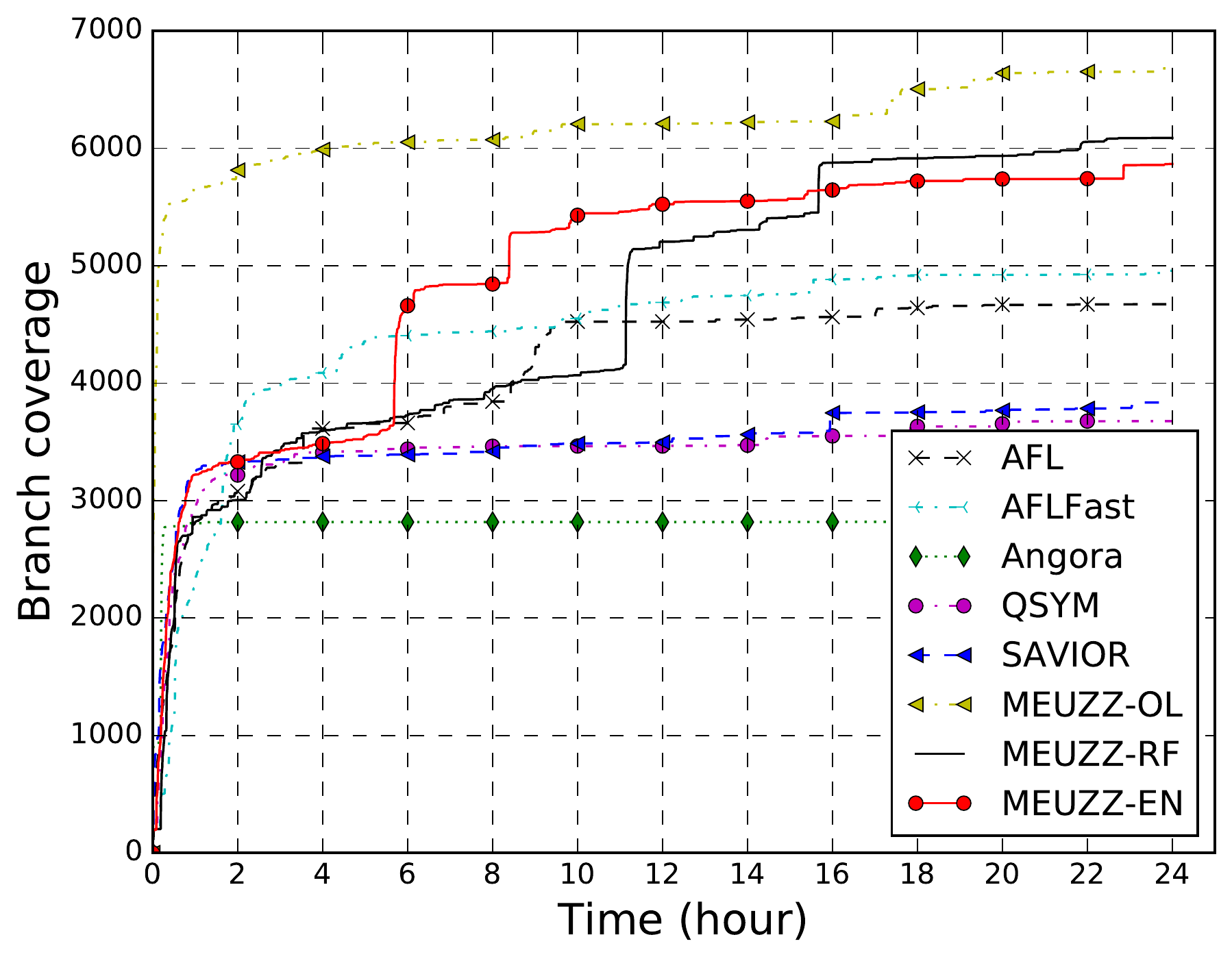}
        \vspace*{-15pt}
        \caption{\scriptsize{tiff2ps branch coverage ($p_{1}$=$6.04*e^{-4}$, $p_{2}$=$0.012$, $p_{3}$=$5.6*e^{-3}$)}}
        \label{fig:cov2:tiff2ps}
    \end{subfigure}  
    \begin{subfigure}[b]{0.23\textwidth}
        \centering
        \includegraphics[width=1\textwidth]{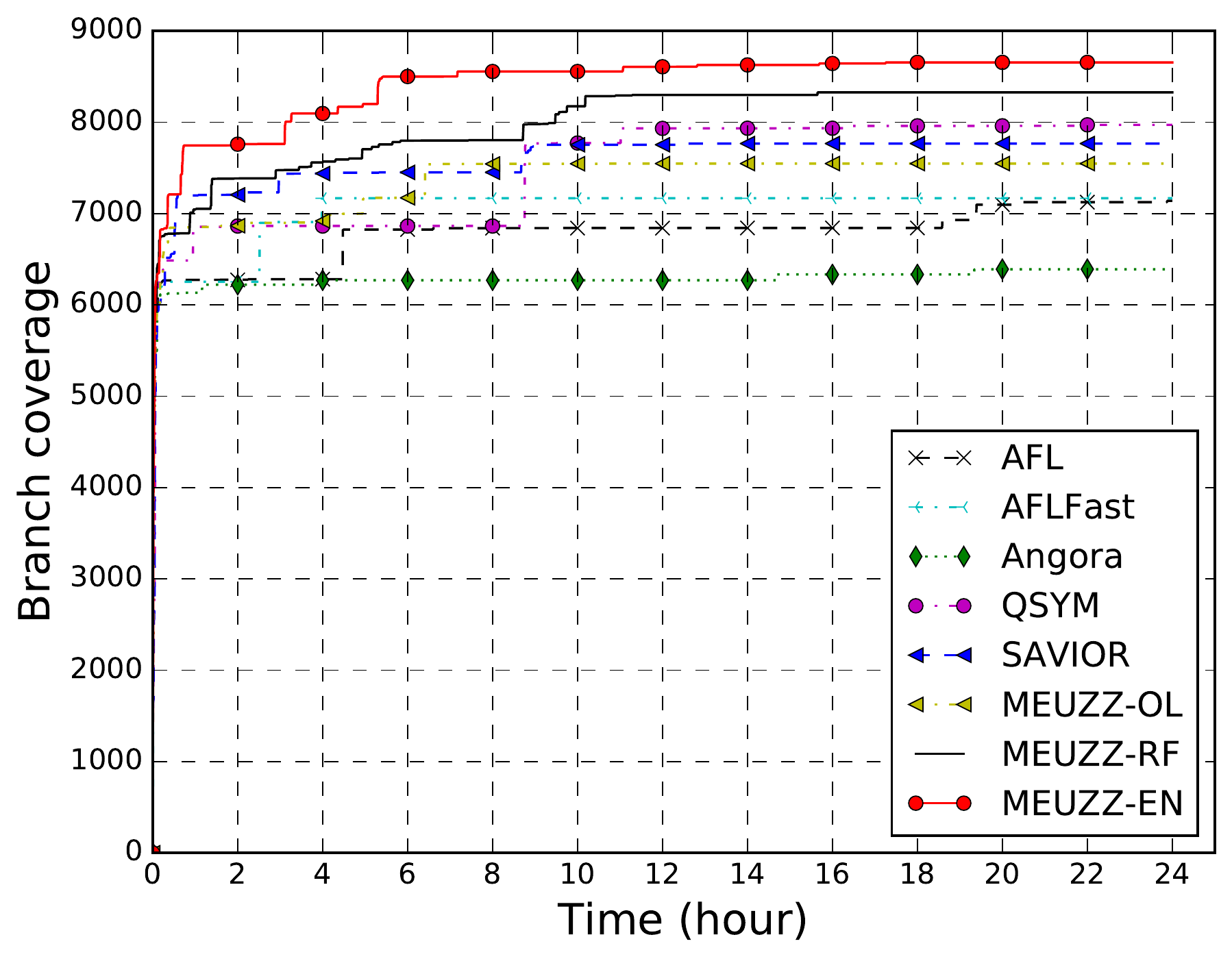}
       \vspace*{-15pt}
        \caption{\scriptsize{jasper branch coverage ($p_{1}$=$0.264$, $p_{2}$=$0.0268$, $p_{3}$=$1.3*e^{-3}$)}}
        \label{fig:cov2:jasper}
    \end{subfigure}\\
    \begin{subfigure}[b]{0.23\textwidth}
        \centering
        \includegraphics[width=1\textwidth]{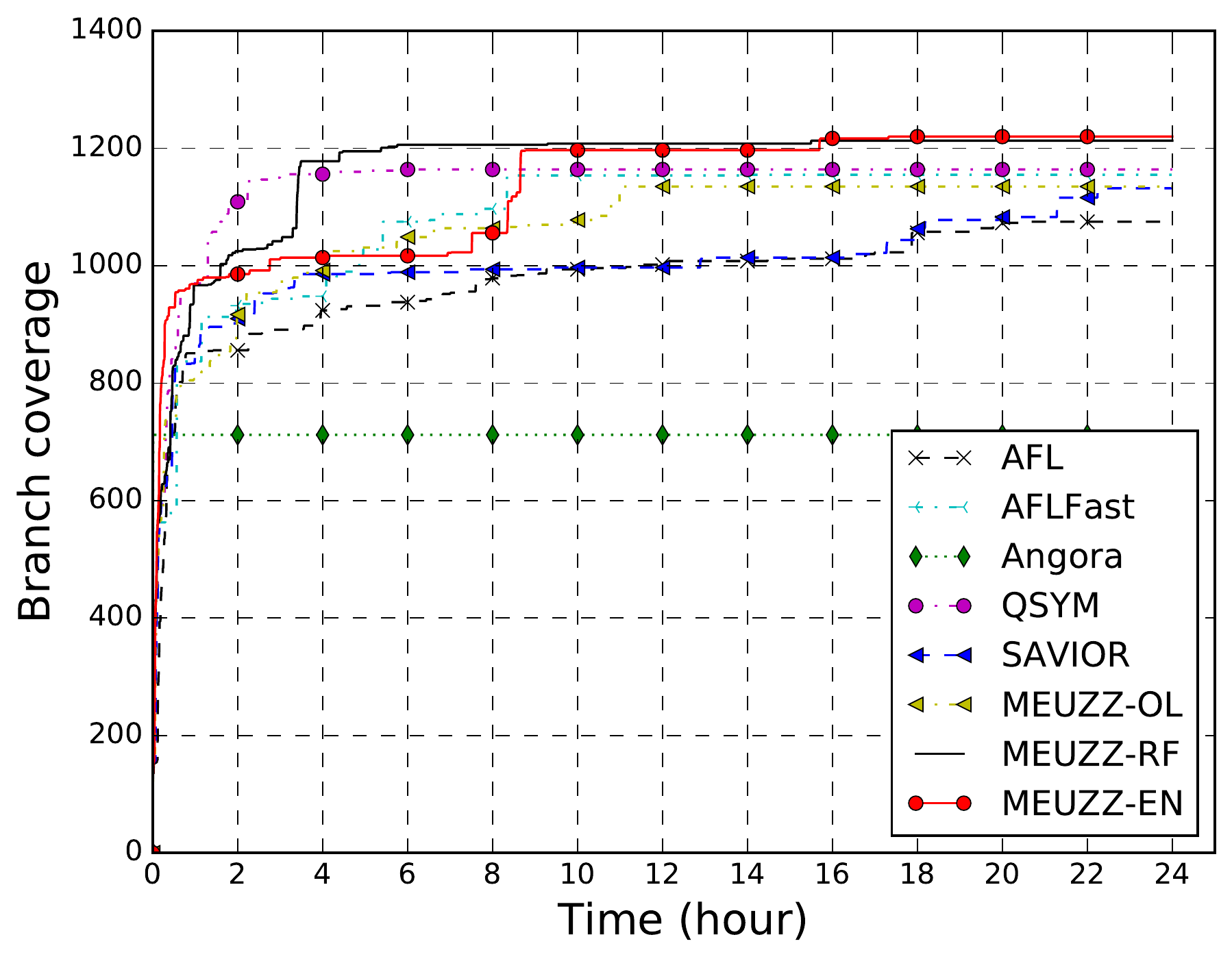}
       \vspace*{-15pt}
        \caption{\scriptsize{readelf branch coverage ($p_{1}$=$0.03$, $p_{2}$=$0.072$, $p_{3}$=$0.037$)}}
        \label{fig:cov2:readelf}
    \end{subfigure}
    \begin{subfigure}[b]{0.23\textwidth}
        \centering
        \includegraphics[width=1\textwidth]{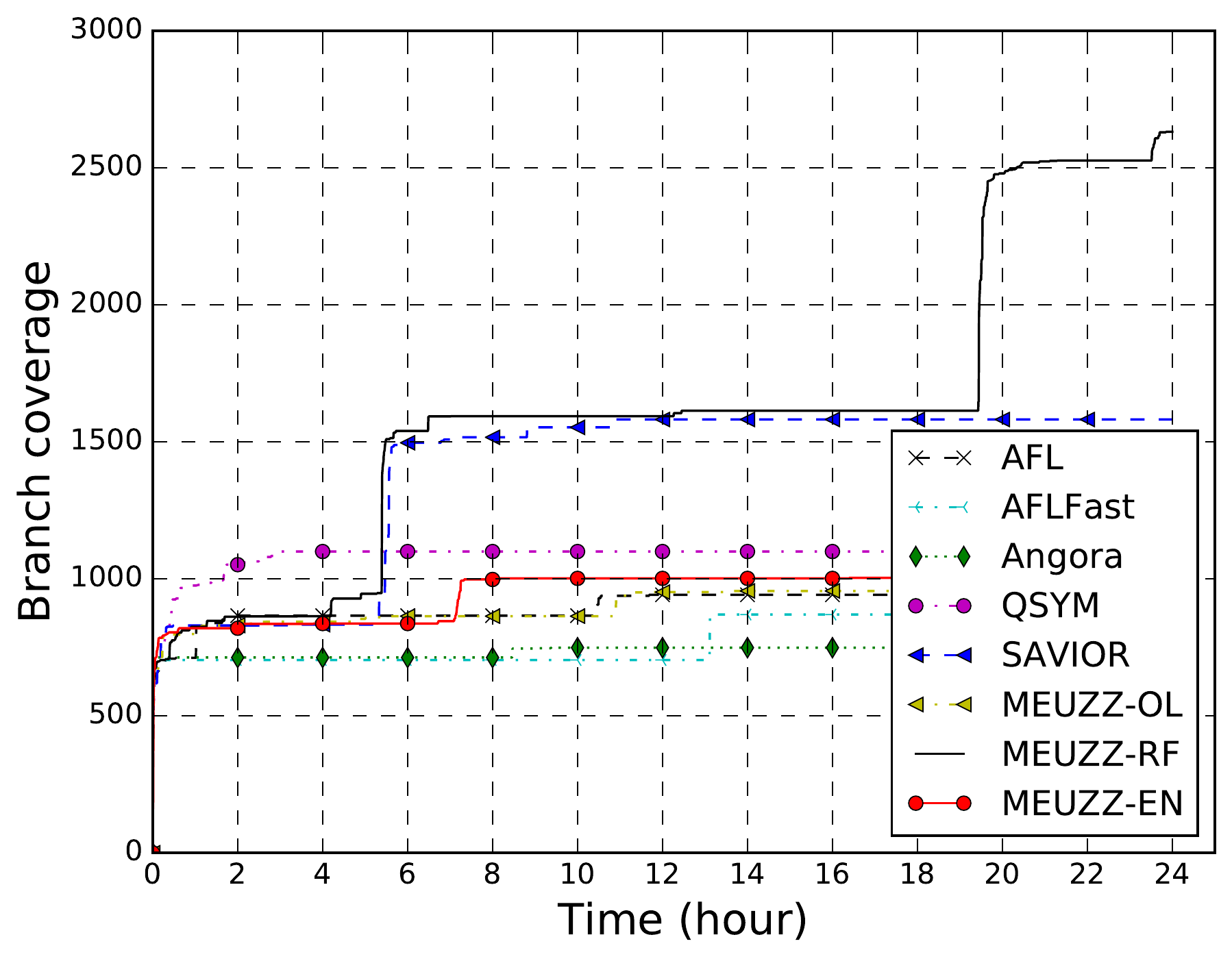}
       \vspace*{-15pt}
        \caption{\scriptsize{djpeg branch coverage ($p_{1}$=$6.04*e^{-3}$, $p_{2}$=$0.012$, $p_{3}$=$3.68*e^{-3}$)}}
        \label{fig:cov2:djpeg}
    \end{subfigure}
    \caption{Branch coverage fuzzing with naive seeds (higher is better). $p_{1}$, $p_{2}$ and $p_{3}$ are p-values in Mann-Whitney U Test by comparing QSYM with MEUZZ-OL, MEUZZ-RF and MEUZZ-EN, respectively.}
    \vspace{-2ex}
    \label{fig:emptycov}
\end{figure}
\setlength{\belowcaptionskip}{-4pt}

Building machine learning models is a valuable but time-consuming task. It is reasonable to build and reuse models where possible.
By reusing a model, one can improve generalization, speed up training, as well as improving the model accuracy.
Also, reusability can be good evidence that our model correctly captured what kind of inputs have higher utility 
when testing the target programs. Hence, we test the reusability of the learned models obtained via the previous fuzzing experiments. 

We conduct an experiment in which we use a pre-trained model for fuzzing the same target program and 
compare the coverage difference. 
We make the following two changes in the experiment performed in \S~\ref{eval:effectiveness}: 
(\rom{1}) the initial seeds are replaced by a naive input that only consists 4 whitespaces; and
(\rom{2}) all \meuzz variants are initialized with the models they learned in the effectiveness test (with valid initial seeds).

\autoref{fig:emptycov} shows the coverage result with Mann-Whitney U Test. 
There are several interesting observations. The most important one is that the \meuzz variants start performing
well even at the beginning of fuzzing compared with when there is no model initialization. 
We believe this improvement is brought by the initial models.
Additionally, ``pure-AFL'' fuzzers do not perform well with this naive initial seed. 
For instance, 
in {\tt tcpdump}, AFL and AFLFast only generate 6 inputs in total after 24 hours of fuzzing
 (see \autoref{fig:cov2:tcpdump:naive}). 
On the contrary, systems augmented with other input generation techniques such as concolic execution and 
taint analysis can generate more inputs and consequently can explore significantly more code. 
Lastly, \meuzz-RF outperforms its peers in 
{\tt djpeg}, and its p-value indicates the improvement is significant ($<0.05$), 
suggesting the non-linear model works better on {\tt djpeg}.

\subsection{Model Transferability}
\label{eval:transferability}
\begin{figure}[ht]
\centering
\includegraphics[scale=0.4]{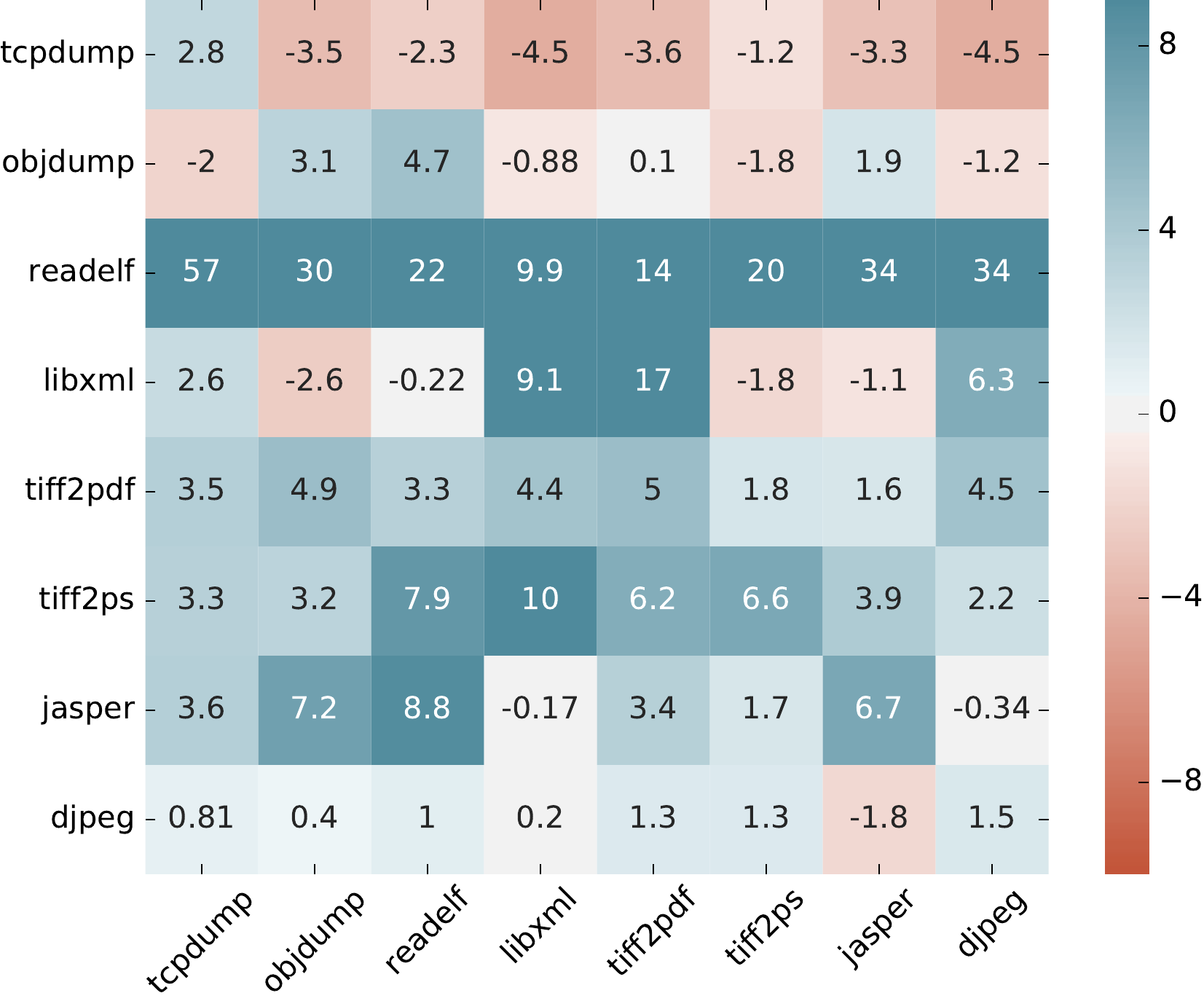}
\caption{This heat map shows Coverage improvement with model initialization for \meuzz-OL over vanilla \meuzz-OL. Y-axis is the tested programs, X-axis is the models used for initialization. Each cell shows the relative coverage comparison (\%). The diagonal values show the coverage improvement on each program after initializing \meuzz with model learn from the same program (reusability). Model transferability is shown in 7 out of the 8 programs.}
\label{fig:trans}
\end{figure}
To further evaluate the model reusability explained in the previous section, we conduct a cross-program experiment 
to determine whether a model trained on one type of program will transfer well to fuzzing a new program. 
This is known as transfer learning in the ML field \cite{transferlearning}.
As far as we know, no prior research has attempted to show this invaluable analysis in fuzzing\cite{sandiasurvey}. 

In this experiment, we augment \meuzz with a pre-trained model from one program and 
compare the result of the fuzzer on different programs with a baseline.
Our baseline is the coverage result from the 
learning effectiveness experiment (\S~\ref{eval:effectiveness}), in which we use valid seeds to bootstrap fuzzing without model initialization.
We choose \meuzz-OL as the representative of our system to measure this transferability experiment.
We then fuzz each program using \meuzz-OL initialized with the 8 pre-learned models; 
the models are fixed afterward.

\autoref{fig:trans} visualizes the comparative coverage improvements (\ie percentage) produced
by each fuzzing configuration. The Y-axis shows the tested program and the X-axis shows the programs 
by which the models are built. This result shows three interesting findings.

First, \meuzz-OL observes 7.1\% more code coverage on average when it is tested on the same program it is initialized with. 
The amount of improvement for each program is shown in the diagonal of \autoref{fig:trans}, from top left to bottom right.
Note that these models are only learned in 24 hours from previous experiments; we expect to see more 
improvement in continuous fuzzing services (\eg \cite{ossfuzz}). 
This again confirms that the previously learned models are reusable.

Second, \meuzz-OL observes improvement in 38 out of 56 cross-testing cases, which shows 67.9\% 
success rate when the model is transferred from a program to another program. 
Among them, 10 cases see more than 10\% coverage improvement. 
Such improvement also indicates that the program-agnostic requirement is satisfied in \meuzz.

Last but not least, we notice different programs have different ``sensitivity'' towards the transferred models. 
For instance, almost all the transferred models can strengthen fuzzing {\tt readelf}, {\tt tiff2pdf}, {\tt tiff2ps} and {\tt djpeg} programs, 
among which {\tt readelf} sees the highest improvement. Interestingly, {\tt readelf} achieves even higher improvement 
when using the {\tt tcpdump} model than the {\tt readelf} model by itself. However,
other programs are only partially accepting foreign models. 
For instance, the model of {\tt tcpdump} can outperform almost all of the programs, 
while none of the other seven external models can improve its fuzzing yields.

Two main reasons can justify the aforementioned observation, namely the number of data points as well as feature importance. 
When there is more data, the model can better generalize \cite{Halevy:2009:UED}.
For instance, the {\tt tcpdump} model 
contains a higher number of seeds compared with others (see \autoref{fig:motiv}), 
which justifies the effectiveness of the transferred model built from the {\tt tcpdump} program.
We also compared the importance of the features of each program (see Appendix~\ref{sec:appendix:feature}).
The shape of the final importance chart in {\tt tcpdump} diverges more from the rest of the programs. Moreover, the values of some features 
such as \textit{Indirect Call} and \textit{Path Length} are higher than other programs.
By looking at these statistics as well as checking the source code of {\tt tcpdump} we noticed {\tt tcpdump} is designed with heavier use of function 
handlers for different types of network packets and recursive loops for parsing packet fields. 
While other models contain different feature value distribution as well as fewer data points, 
which justify the failure of using them to improve fuzzing {\tt tcpdump}.

\subsection{Discovered Bugs}
\label{eval:bugs}

\begin{table}
 \caption {The table shows the unique bugs found by all evaluated fuzzers.}
 \centering
 \resizebox{0.47\textwidth}{!}{
     \begin{tabular}{l|cccccc|c}
     \toprule
     Program & AFL & AFLFast & Angora & QSYM & Savior & \meuzz & All unique bugs \\
     \hhline{========}
          tcpdump     & 14 &  13   & 12 &  11 & 12  & 14 & 14\\ 
          objdump     & 2  &   2   &  5 &  2  & 8   & 6  & 9\\ 
          readelf     & 3  &   2   &  5 &  5  & 4   & 4  & 6\\ 
          tiff2pdf    & 1  &   1   &  1 &  2  & 2   & 2  & 2\\ 
          tiff2ps     & 1  &   2   &  2 &  5  & 4   & 6  & 6\\ 
          jasper      & 2  &   1   &  0 &  3  & 1   & 6  & 8\\ 
          djpeg       & 9  &   7   &  7 &  9  & 9   & 9  & 9\\ 
     \hhline{========}
         Total 		  & 32 &  28   &  32  &  37 & 40  & 47 & 54 \\ 
     \bottomrule
     \end{tabular}
     }
     \label{tab:bugcompare}
\end{table}

To prove the effectiveness of our system in discovering new bugs, we performed various analyses. 
We manually analyzed all of the reported undefined behaviors and crashes. 
UBSan reports a large
amount of undefined behaviors; however, the majority of them are deemed benign after our triage process.
We also triage additional bugs with the help of ASAN \cite{AddressSanitizer} and LeakSAN \cite{LeakSanitizer}. 

Table~\ref{tab:bugcompare} shows our triage result for all the fuzzers. In total, 54 unique
bugs were uncovered. \meuzz outperforms other fuzzers and found 47 unique bugs, which supports the fact that higher code coverage correlates to a higher number of triaged bugs. Due to space
limit, we present more detailed triage result and one of the discovered bugs only found by \meuzz in Appendix~\ref{sec:appendix:meuzzbug}. 
This 
result shows \meuzz is more effective in terms of finding bugs than 
state-of-the-art systems with manually crafting heuristics.

\section{Related Work}
\label{sec:related}

\subsection{ML for Fuzzing}
Despite the promising potential to improve fuzzing, the application of ML has not been very-well investigated in the past and 
only a few research have leveraged ML. ML can be integrated into various stages of fuzzing, from input generation to crash categorization. 

\smallskip
\noindent
\textbf{Input generation:}
The most intelligent stage of fuzzing has been
the input generation stage so far, thanks to genetic algorithms.
Deep learning (DL) techniques have been recently applied to input generation for both mutation/generation-based fuzzing. 
Such approaches \cite{DLIF,neuralfuzzing,neuzz} use various neural network methods to learn the patterns that exist in input files and 
then identify the likely input forms to trigger new coverage. 
Similarly, reinforcement learning (RL) \cite{RLfuzzing} can learn input grammar for generation-based fuzzers.

\smallskip
\noindent
\textbf{Crash analysis:}
Automating the analysis of outputs/crashes generated by fuzzers is another ML application. 
For instance, ML can be used to categorize crashes by identifying the root cause of them. 
This helps remove duplicate outputs and therefore reduces manual analysis effort \cite{ReBucket}. 
Or another example is employing ML to predict whether the reported crashes by fuzzers are exploitable \cite{ExploitMeter}.

To the best of our knowledge, there has not been any research that practices ML for seed selection.
In general, the practicality of ML for fuzzing has not been shown clearly in the past due to the 
uncertainty about reusability and transferability.

\subsection{Seed Scheduling Heuristics}
\smallskip
\noindent
\textbf{Scheduling in fuzzing:}
FuzzSim\cite{fuzzingscheduling} models the seed scheduling problem as a weighted 
coupon collector problem and found out that scheduling can have a direct impact on fuzzing campaign yields. 
Later, in grey-box fuzzing, AFL\cite{afl} implements a scheduling algorithm that consists of
simple heuristics such as preferring first seed with new coverage, 
and with smaller size and less execution time.
This simple algorithm is later improved by Fairfuzz\cite{fairfuzz} and AFLFast\cite{aflfast} which steer the fuzzer towards less explored paths.

\smallskip
\noindent
\textbf{Scheduling in hybrid testing:}
As hybrid testing becomes more popular, seed scheduling also becomes a research topic. Driller\cite{driller} implements a random scheduling algorithm, while
QSYM\cite{qsymimpl} implements heuristics similar to AFL.
Later, DigFuzz\cite{digfuzz} shows the ineffectiveness of random scheduling and proposes a Monte-Carlo model to predict the difficulty of each path explored
by the fuzzer by far, and send the most difficult ones to concolic executor. 
SAVIOR\cite{savior}, on the other hand, uses bug-driven scheduling heuristics. By selecting the seeds that can reach more sanitizer instrumentations, it triggers more bugs in the given timeframe than other fuzzers.

Compared with these approaches, \meuzz applies machine learning techniques that can learn a utility prediction model, which is adaptive to the program being tested. As our evaluation suggests, this approach is more scalable and more performant than the manual-crafting scheduling heuristics.

\section{Conclusion}
\label{sec:conc}
We present \meuzz, a hybrid fuzzing system featuring machine learning and data-driven seed scheduling.
Theoretically, \meuzz is more generalized than systems using fixed seed selection heuristics. 
For effective integration of machine learning workloads into the online hybrid fuzzing loop, 
\meuzz follows the requirements of being utility relevant, online friendly and program agnostic for its feature engineering and label inference.
Our evaluation shows that \meuzz outperforms state-of-the-art fuzzers in both code coverage and bug discovery. 
In addition, the learned models demonstrate good reusability and transferability, making it more practical to apply machine learning to hybrid fuzzing.

\section*{Acknowledgment}

The authors would like to thank the anonymous reviewers for their insightful
comments. This project was supported by 
the National Science Foundation (Grant\#: CNS-1748334) %
and the Office of Naval Research (Grant\#: N00014-17-1-2891). %
Any opinions, findings, and conclusions or recommendations expressed in this
paper are those of the authors and do not necessarily reflect the views of the
funding agencies.

\bibliographystyle{plain}
\bibliography{ref}
\begin{appendices}

\section{Why use UBSAN}
\label{sec:appendix:sanitizerchoice}
Note that although the design of MEUZZ is generically compatible with mainstream sanitizers~\cite{asan,ubsanlist,msan}, our implementation uses UBSan for the following
reasons: (\rom{1}) UBSan instruments programs with pure static checks that can be easily
converted to solvable SMT constraints. In contrast, other sanitizers, such as ASAN and MSAN, employ red-zones and status bitmap, which are less amenable to constraint solving.  
(\rom{2}) Our concolic engine is based on SAVIOR's KLEE, which uses UBSan as the primary sanitizer. Using UBSan makes
concolic execution more effective as shown in~\cite{savior}.

\section{Bugs found by \meuzz}
\label{sec:appendix:meuzzbug}

We provide a more detailed triage information of the bugs found by \meuzz.
In total, \meuzz found 30 undefined behaviors, among which 21 have been confirmed/fixed so far by the developers and the rest are pending. For the reported bugs, we found
the potential UBs with UBSAN\cite{ubsanlist} and manual analysis; we found the memory errors and DoS
with ASAN\cite{asan} and memory leaks with LeakSAN\cite{LeakSanitizer}.

\begin{table}[h]
 \caption {The table shows the discovered bugs by \meuzz. UB, ME, DoS, and ML refers to Undefined Behavior, Memory Error, 
 Denial of Service, and Memory Leak, respectively.}
 \resizebox{0.49\textwidth}{!}{
     \begin{tabular}{l|cccc|c}
     \toprule
     Program & Potential UB & ME & DoS & ML & Confirmed \\
     \hhline{======}
          tcpdump     & 14 &    &   &    & 4    \\
          objdump     & 4  &     &  &  2 &   \\
          readelf     & 2  & 1    &  &  1 & 1   \\
          tiff2pdf    &  &      & 2 &   & 2    \\
          tiff2ps     & 1  & 4    & 1 &    & 4   \\
          jasper      &    &     & 4 & 2   & 4    \\
          djpeg       & 9 &     &  &    & 6     \\
     \hhline{======}
         Total 		  & 30 & 5   &  7 &  5  & 21     \\
     \bottomrule
     \end{tabular}
     }
     \label{tab:discoveredbug}
\end{table}

\begin{figure}[ht]
\includegraphics[scale=0.9]{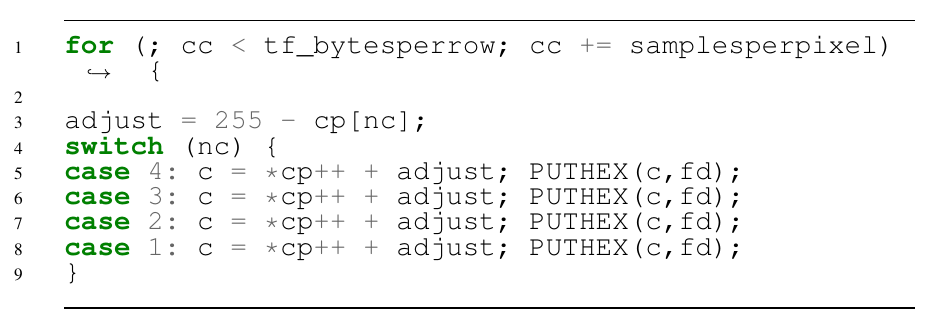}
\caption{Off-by-one heap read overflow in tiff2ps.}
\label{fig:tiffbug}
\end{figure}

One of the heap overflow vulnerabilities in {\tt tiff2ps}
is discovered only by \meuzz. \autoref{fig:tiffbug} shows the vulnerable code snippet. 
This bug has been confirmed and fixed by the developers. It is an out-of-bound read vulnerability that can lead to information disclosure. 
The vulnerability takes place at {\tt PSDataColorContig} function where {\tt cp} buffer with the size of 4 bytes is allocated in heap and the 5th element of the buffer is accessed by {\tt cp[4]} which leads to out-of-bound read. To trigger this bug, the loop needs to be executed without early breaks. Moreover, to control the buffer size, the input needs to satisfy many constraints in the {\tt TIFFScanlineSize} function so that it will return value 4.
Based on the feature importance of {\tt tiff2ps} (Appendix~\ref{sec:appendix:feature}), \emph{Size}, \emph{Cmp} and \emph{External Call} play more important roles in its model, we believe this is why \meuzz is able to guide the fuzzer to explore and trigger this bug. On the contrary, by replaying the fuzzing corpora, we found that other fuzzers miss this bug because they either exit the loop early or fail the checks in {\tt TIFFScanlineSize}.

\section{Discussion on Extra Experiments}
\label{sec:appendix:other}

We attempted to compare \meuzz with many state-of-the-art fuzzing systems but cannot conduct an apple-to-apple comparison with some of them due to various reasons.

Driller uses \cite{angrtrac7:online} as its concolic engine, which has limited support for system calls,
causing the engine's failure to generate new test cases. Similar issue was also reported by Insu at el.\cite{qsyminsu}.
Vuzzer and T-Fuzz do not have support for concurrent fuzzing. After discussing
with the developers we assigned only one core to them and run them for 72 hours
(3$\times$24) instead of 24 hours.
We report the branch coverage results of Vuzzer and T-Fuzz in Table~\ref{tab:tfuzzvuzzer}.

\begin{table}[ht]
 \caption {The table shows the number of branches covered by Vuzzer and T-Fuzz. \textcolor{red}{\ding{55}} means fuzzer crashed on the program.}
 \centering
 \resizebox{0.3\textwidth}{!}{
     \begin{tabular}{l|c|c}
     \toprule
     Program & Vuzzer& T-Fuzz \\
     \hhline{===}
          tcpdump     & 1103  & 11566      \\
          objdump     & 711     & 4216   \\
          readelf     & 1025      & 842    \\
          libxml2     &   715    &  \textcolor{red}{\ding{55}}   \\
          tiff2pdf    &   \textcolor{red}{\ding{55}}     & 4892   \\
          tiff2ps     &  \textcolor{red}{\ding{55}}    & 3534   \\
          jasper      &  \textcolor{red}{\ding{55}}    & 6084      \\
          djpeg       &  1317   & 763      \\
     \bottomrule
     \end{tabular}
     }
     \label{tab:tfuzzvuzzer}
\end{table}

\section{Detailed Feature Importance Study}
\label{sec:appendix:feature}

Figure~\ref{fig:modelchange} demonstrates how the randomly initialized model evolved with
more and more training data available during fuzzing. \meuzz automatically identifed
which features are more important for each specific programs, showing it is more
scalable than manually-written heuristics.

\setlength{\belowcaptionskip}{3pt}
\begin{figure*}[t]
    \centering   
    \begin{subfigure}[b]{0.24\textwidth}
        \centering
        \includegraphics[width=1\textwidth]{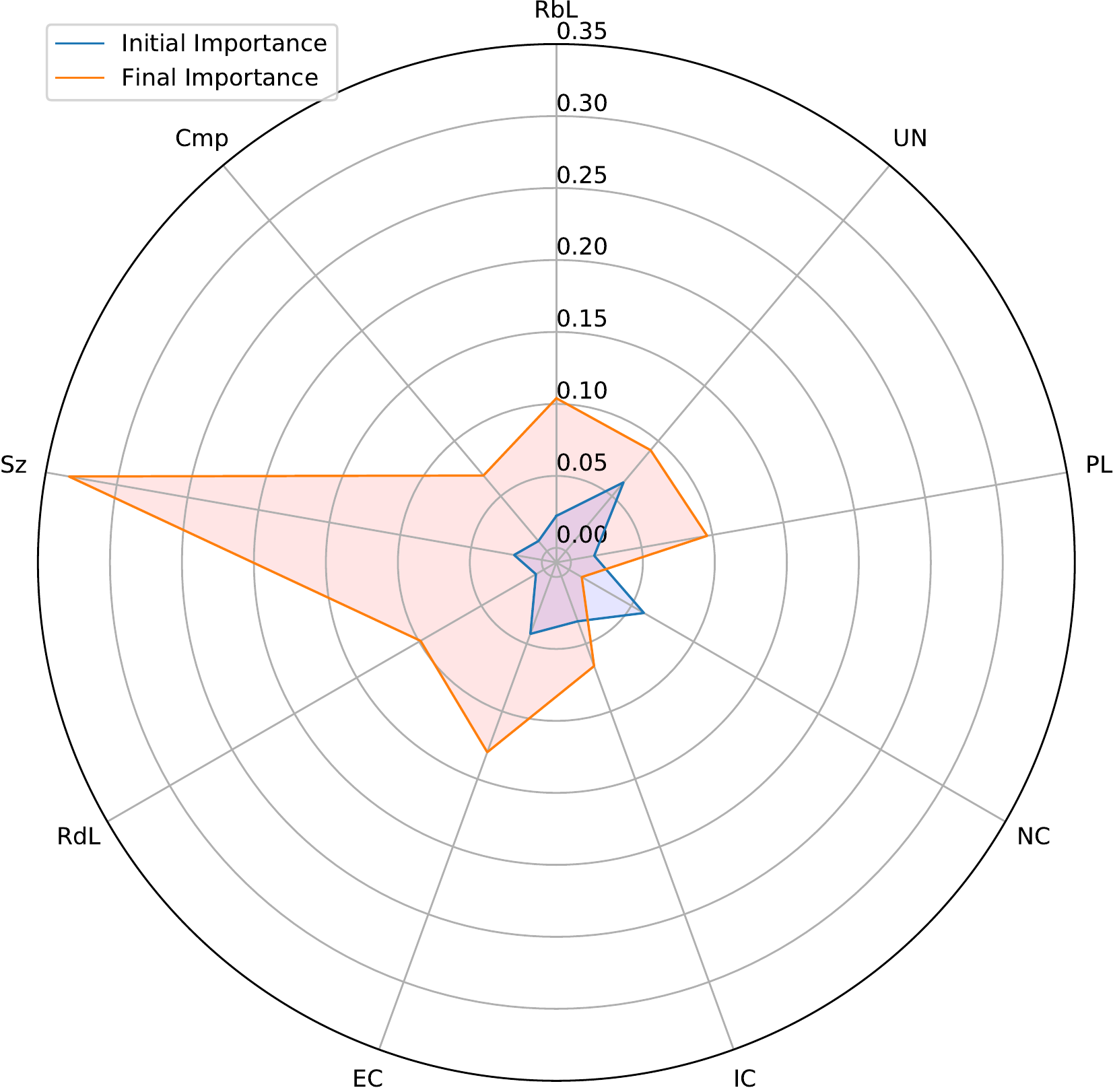}
        \vspace*{-15pt}
        \caption{\scriptsize{Feature importance in tcpdump}}
        \label{fig:feat:tcpdump}
    \end{subfigure}
        \begin{subfigure}[b]{0.24\textwidth}
        \centering
        \includegraphics[width=1\textwidth]{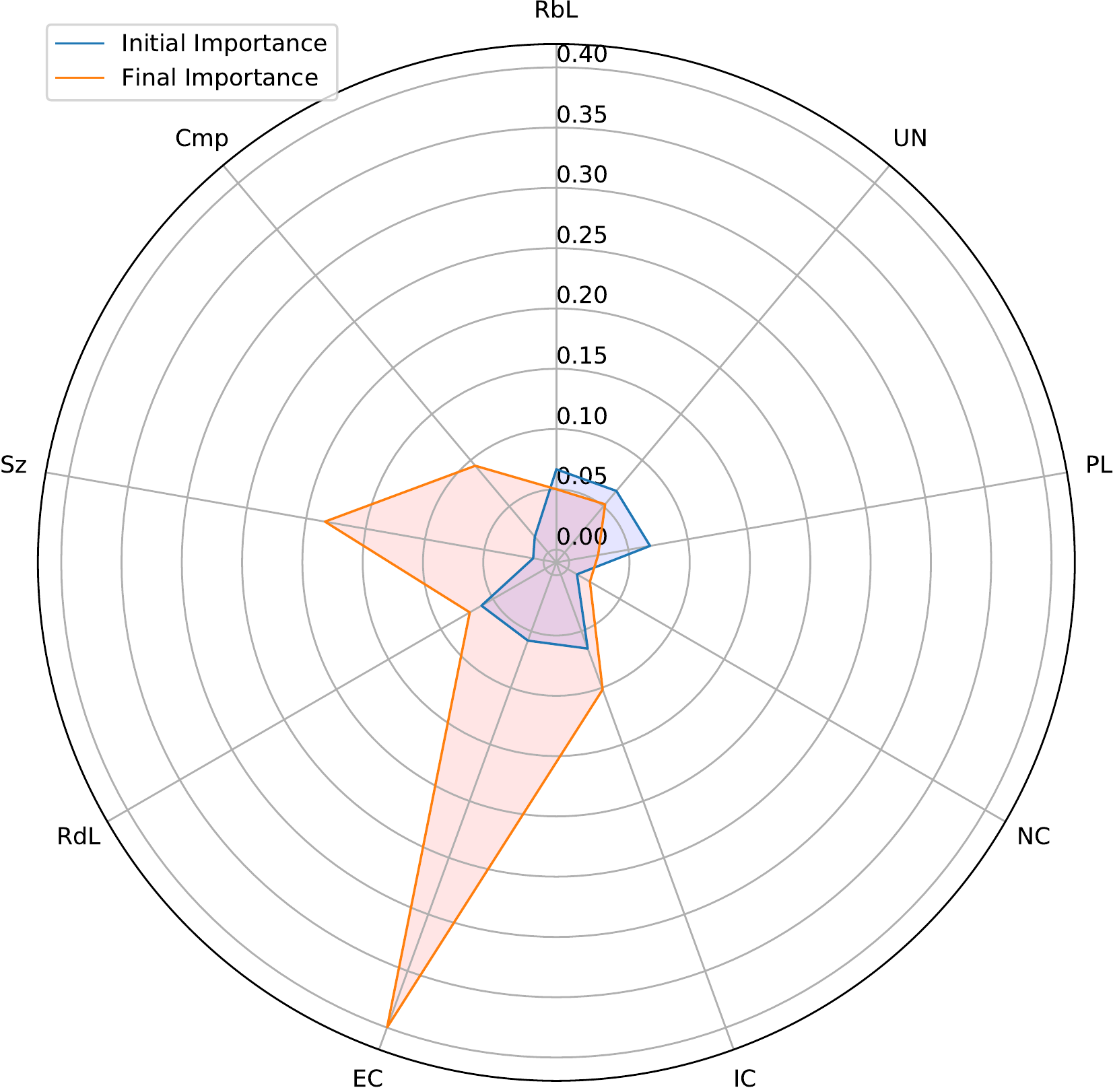}
         \vspace*{-15pt}
        \caption{\scriptsize{Feature importance in objdump}}
        \label{fig:feat:objdump}
    \end{subfigure}
    \begin{subfigure}[b]{0.24\textwidth}
        \centering
        \includegraphics[width=1\textwidth]{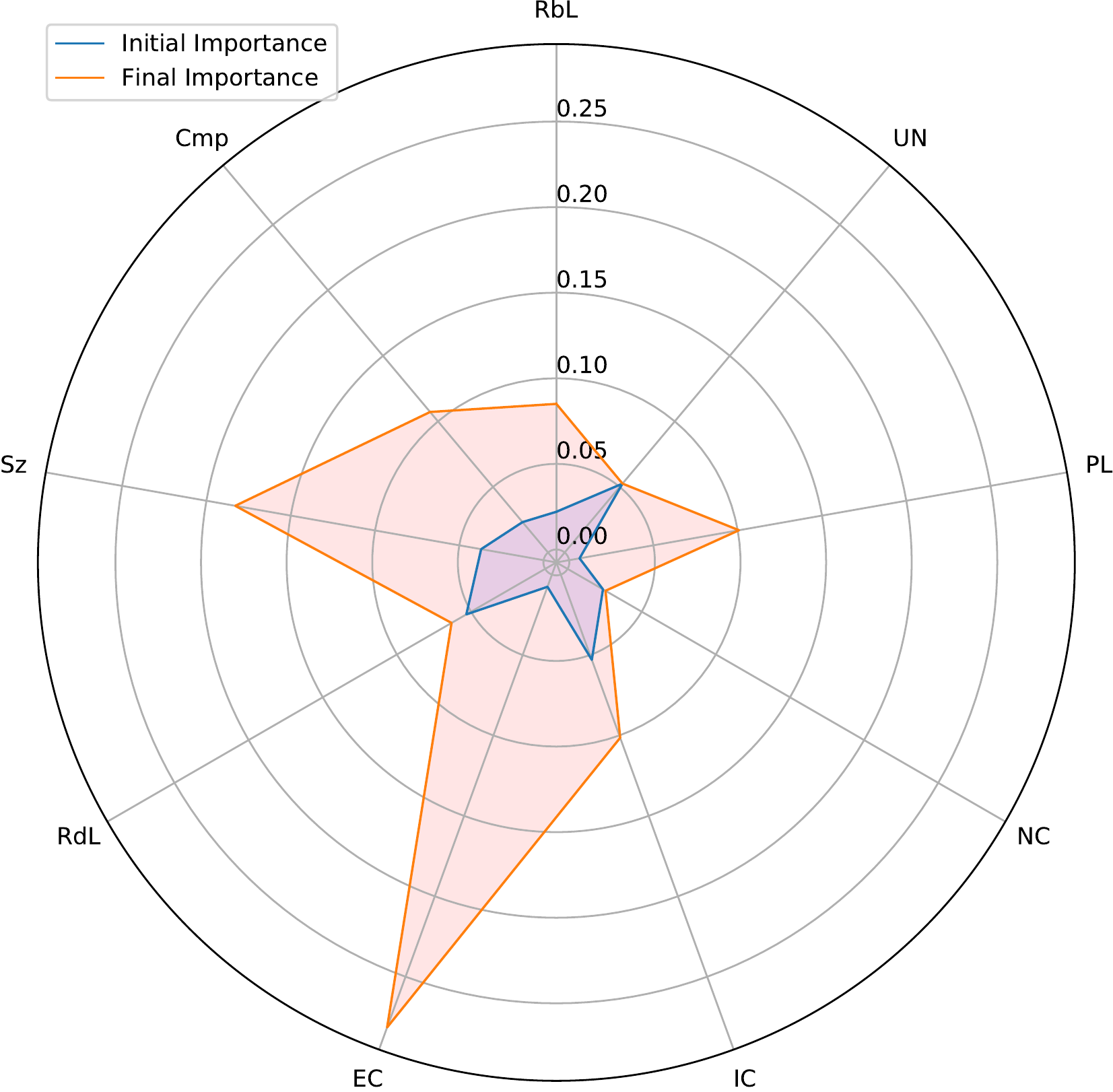}
         \vspace*{-15pt}
        \caption{\scriptsize{Feature importance in libxml}}
        \label{fig:feat:libxml}
    \end{subfigure}
    \begin{subfigure}[b]{0.24\textwidth}
        \centering
        \includegraphics[width=1\textwidth]{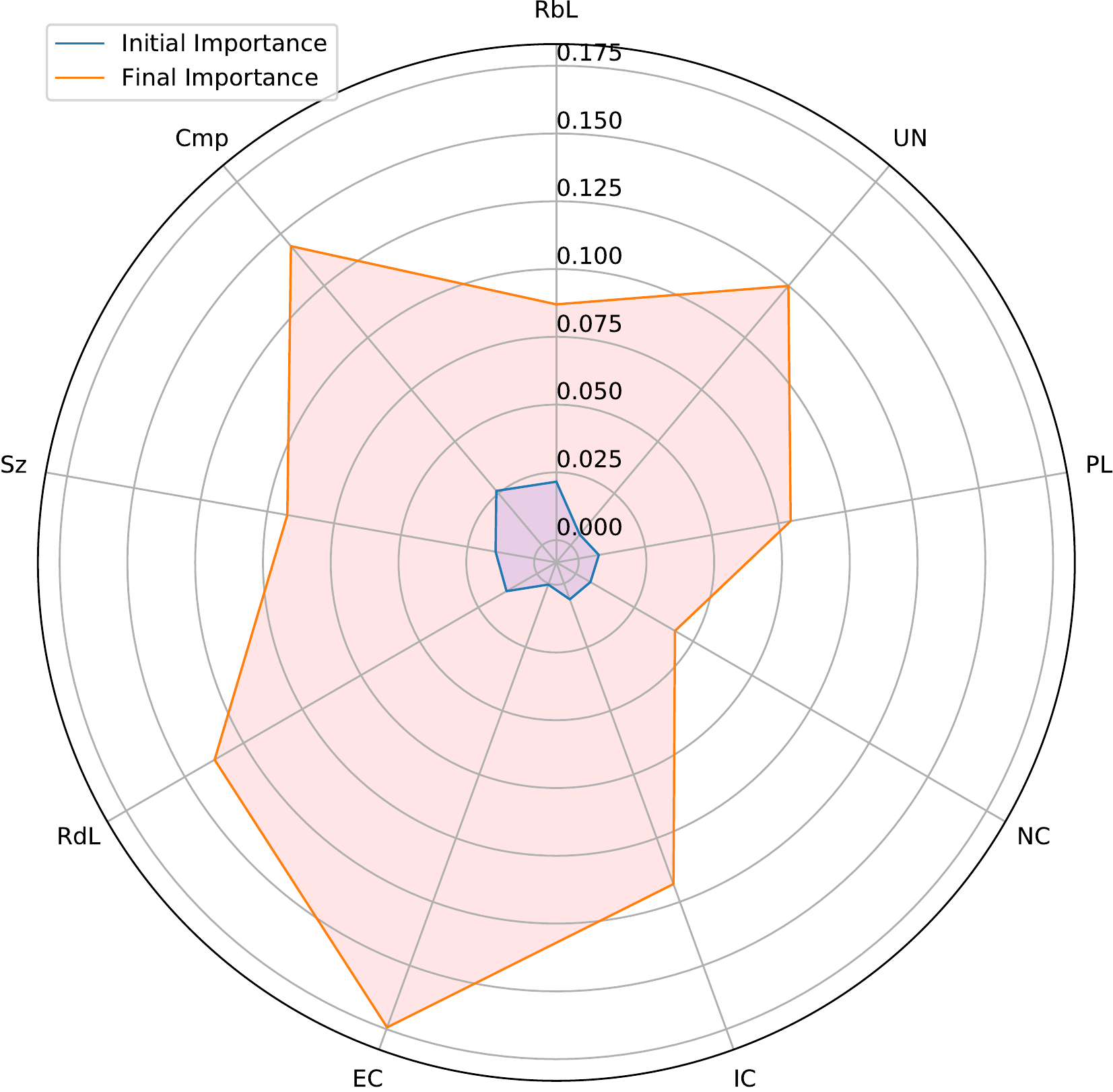}
       \vspace*{-15pt}
        \caption{\scriptsize{Feature importance in tiff2pdf}}
        \label{fig:feat:tiff2pdf}
    \end{subfigure}\\
    \begin{subfigure}[b]{0.24\textwidth}
        \centering
        \includegraphics[width=1\textwidth]{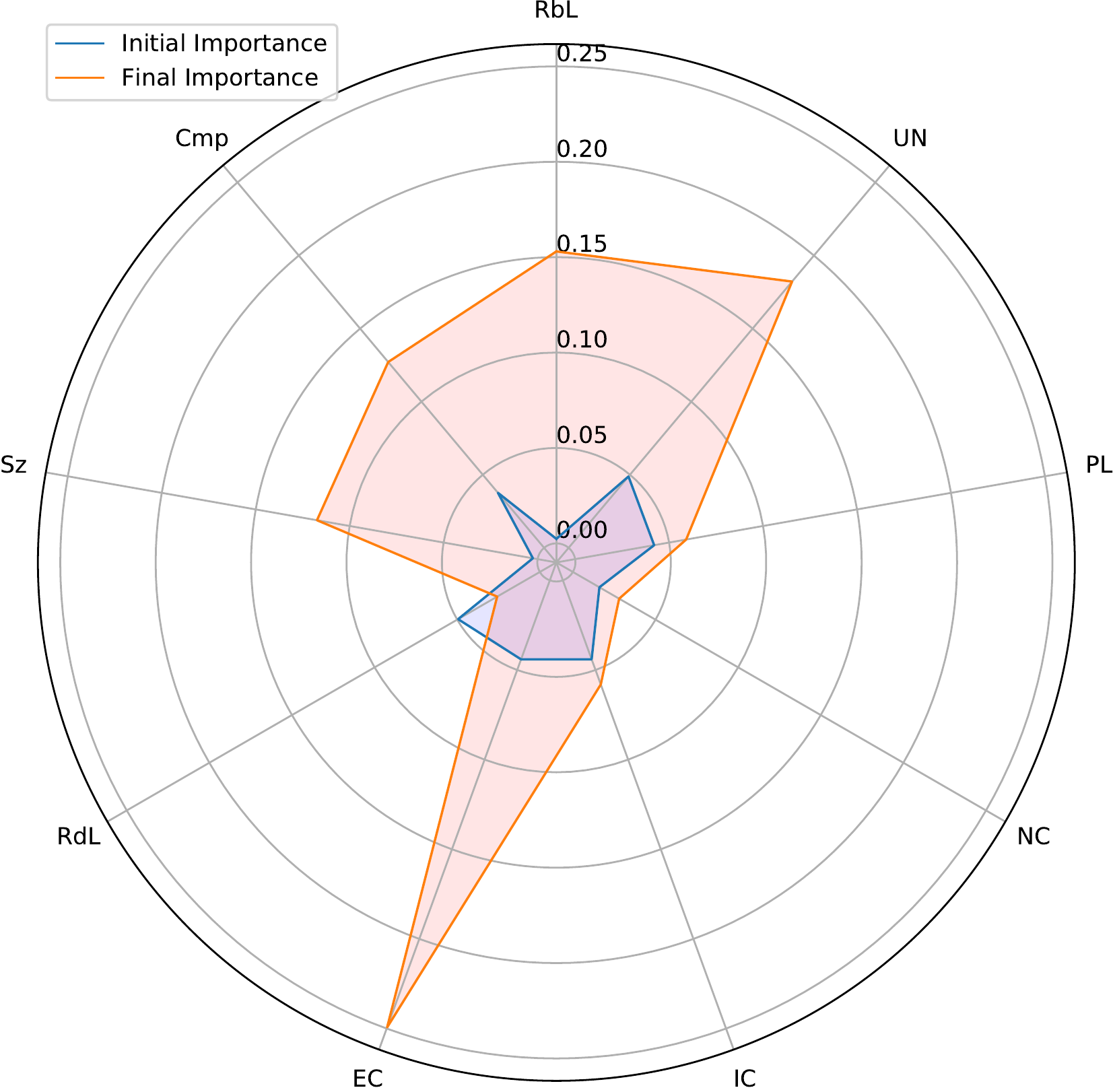}
        \vspace*{-15pt}
        \caption{\scriptsize{Feature importance in tiff2ps}}
        \label{fig:feat:tiff2ps}
    \end{subfigure}  
    \begin{subfigure}[b]{0.24\textwidth}
        \centering
        \includegraphics[width=1\textwidth]{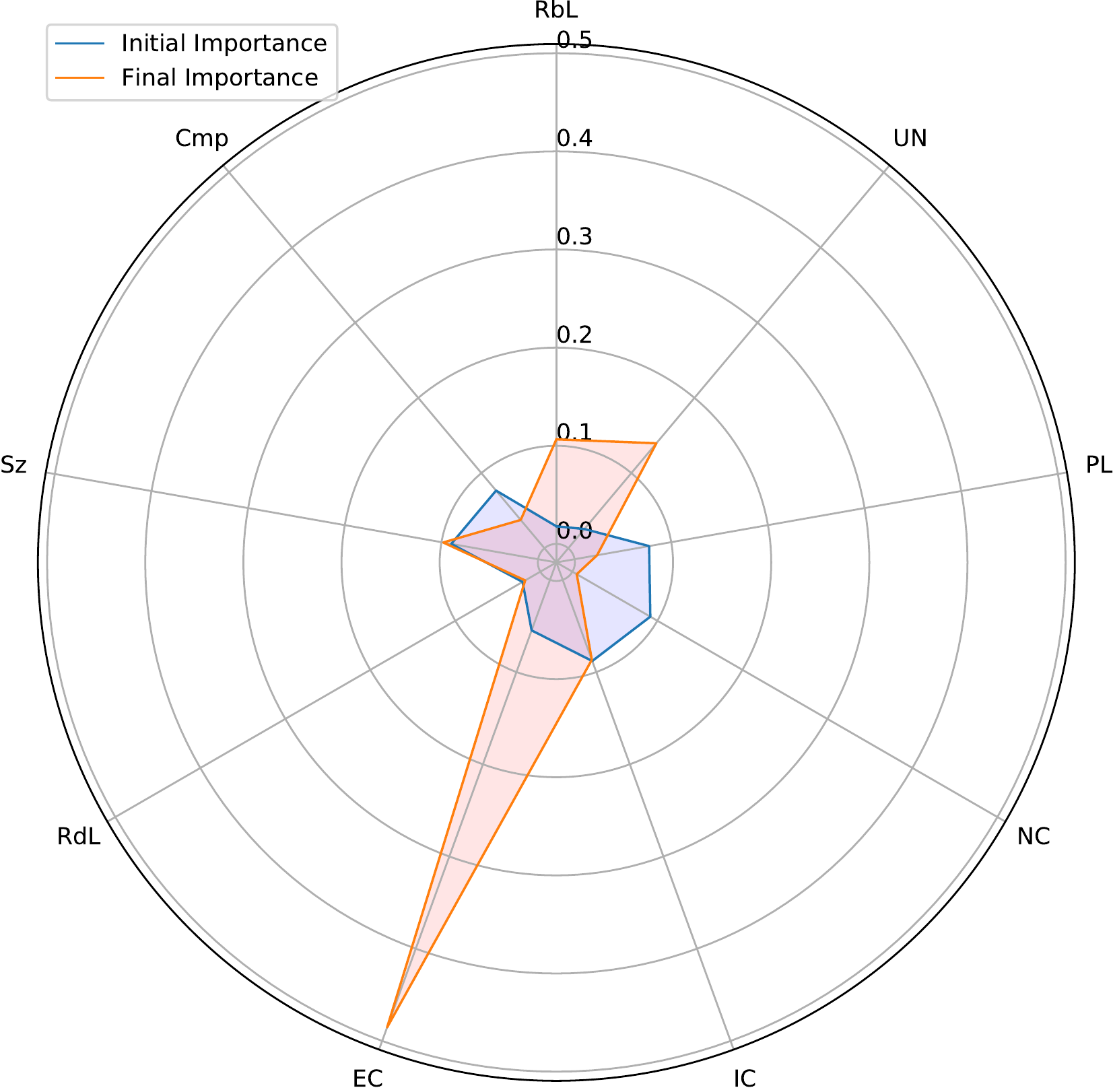}
       \vspace*{-15pt}
        \caption{\scriptsize{Feature importance in jasper}}
        \label{fig:feat:jasper}
    \end{subfigure}
    \begin{subfigure}[b]{0.24\textwidth}
        \centering
        \includegraphics[width=1\textwidth]{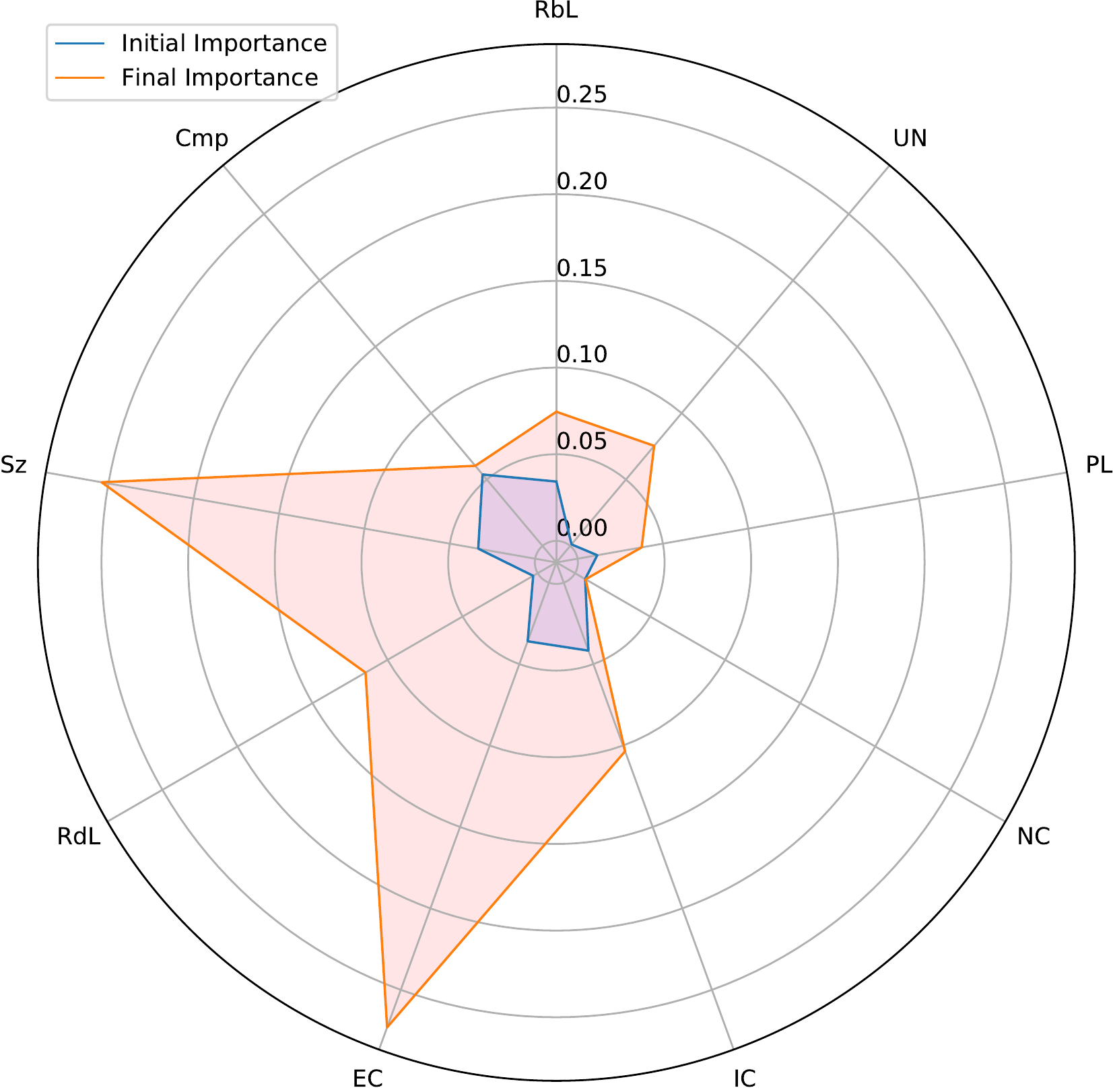}
       \vspace*{-15pt}
        \caption{\scriptsize{Feature importance in readelf}}
        \label{fig:feat:readelf}
    \end{subfigure}
    \begin{subfigure}[b]{0.24\textwidth}
        \centering
        \includegraphics[width=1\textwidth]{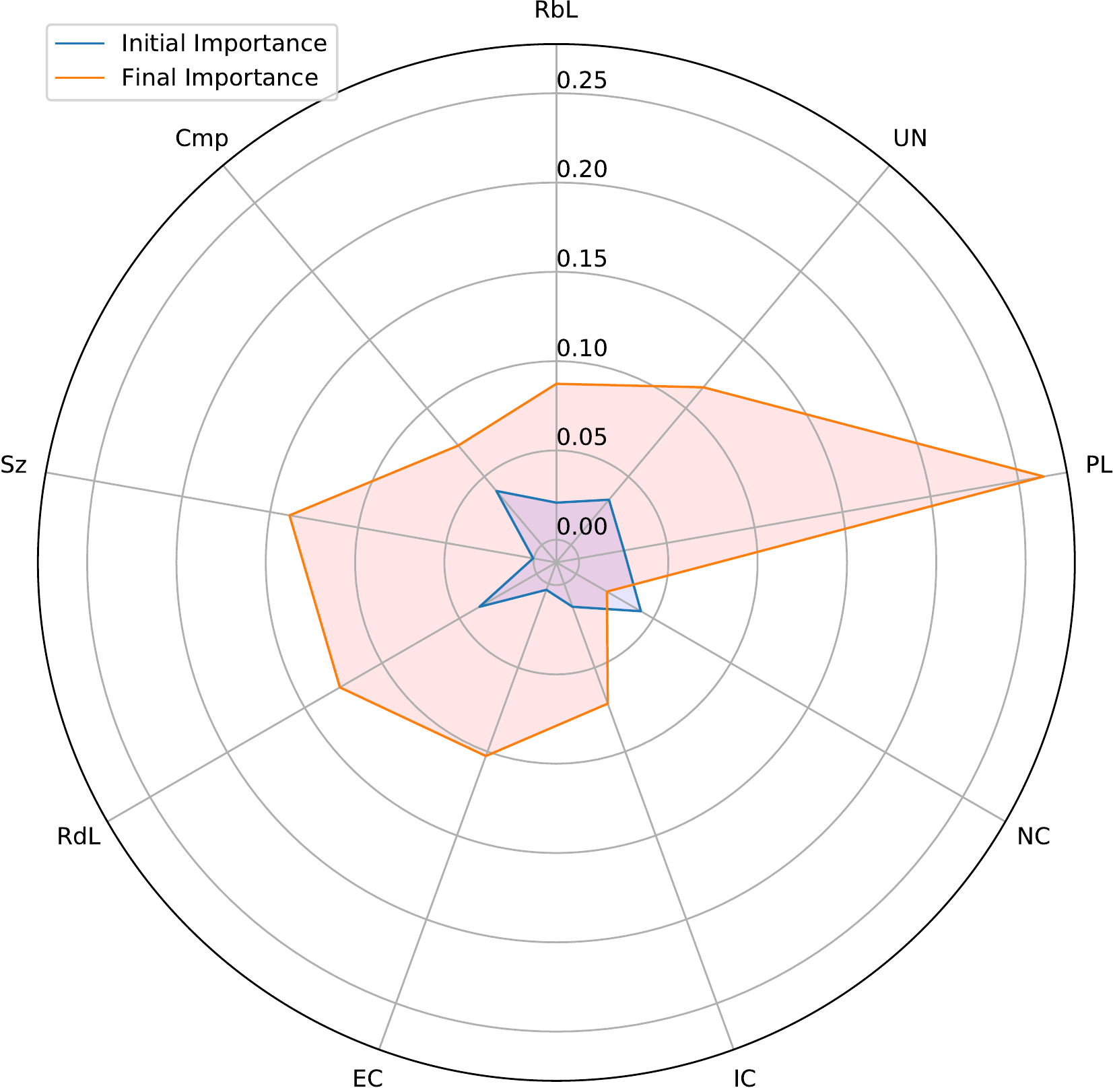}
       \vspace*{-15pt}
        \caption{\scriptsize{Feature importance in djpeg}}
        \label{fig:feat:djpeg}
    \end{subfigure}
    \caption{Feature importance extracted from models learned in the effectiveness test (\S~\ref{eval:effectiveness}). 
    The initial importances are randomly generated. 
    	Sz: \emph{Size}, RdL: \emph{Reached Label}, EC: \emph{External Call}, IC: \emph{Indirect Call}, NC: \emph{New Coverage}, PL: \emph{Path Length}, UN: \emph{Undiscovered Neighbors}, RbL: \emph{Reachable Labels}, Cmp: \emph{Comparisons}.}
    \vspace{-2ex}
    \label{fig:modelchange}
\end{figure*}
\setlength{\belowcaptionskip}{-4pt}

\end{appendices}
\end{document}